\documentclass[journal=jctcce,manuscript=article]{achemso}

\usepackage[utf8]{inputenc}
\usepackage{amsmath, amssymb}

\usepackage[]{siunitx} 
	\DeclareSIUnit\molecule{molecule}
	\DeclareSIUnit\debye{D}
	\DeclareSIUnit\au{a.u.}
	\DeclareSIUnit\Buckingham{B}
        \AtBeginDocument{\RenewCommandCopy\qty\SI}

\usepackage{graphicx}

\graphicspath{{img/}}

\usepackage{afterpage}
\usepackage{amsmath}

\usepackage{hyperref}

\usepackage{color}
\usepackage{braket}
\usepackage{bm}

\newcommand{\ai}{\textit{ab initio}}

\newcommand{\AS}{\ensuremath{A\,{}^{2}\Pi}}
\newcommand{\CS}{\ensuremath{C\,{}^{2}\Pi}}

\newcommand{\BS}{\ensuremath{B\,{}^{2}\Sigma^{+}}}
\newcommand{\DS}{\ensuremath{D\,{}^{2}\Sigma^{+}}}

\newcommand{\Xstate}{\ensuremath{X\,{}^{1}\Pi}}
\newcommand{\Cstate}{\ensuremath{C\,{}^{1}\Sigma^+}}
\newcommand{\Zstate}{\ensuremath{2\,{}^{1}\Sigma^+}}

\newcommand{\cm}{cm$^{-1}$}

\newcommand{\brkteq}[3]{\bra{#1}#2\ket{#3}}

\usepackage{blindtext}
\usepackage{multicol}

\newcommand{\duo}{{\sc Duo}}
\newcommand{\Duo}{{\sc Duo}}

\newcommand{\ts}{\textsuperscript}

\newcommand{\derivBra}[1]{\Big\langle\frac{d #1}{dr}\Big|}
\newcommand{\derivKet}[1]{\Big|\frac{d #1}{dr}\Big\rangle}
\newcommand{\derivBraket}[2]{\Big\langle\frac{d #1}{dr}\Big| \frac{d #2}{dr}\Big\rangle}
\newcommand{\bigBrkt}[2]{\Big\langle #1 \Big| #2 \Big\rangle}
\newcommand{\bigBrktOp}[3]{\Big\langle #1 \Big| #2 \Big| #3 \Big\rangle}
\newcommand{\Ut}{\boldsymbol{U}}
\newcommand{\Utdag}{\boldsymbol{U}^{\dagger}}

\author{Ryan P. Brady}
\affiliation{Department of Physics and Astronomy, University College London, Gower Street, WC1E 6BT London, United Kingdom}

\author{Charlie Drury}
\affiliation{Department of Physics and Astronomy, University College London, Gower Street, WC1E 6BT London, United Kingdom}

\author{Sergei N. Yurchenko}
\affiliation{Department of Physics and Astronomy, University College London, Gower Street, WC1E 6BT London, United Kingdom}
\email{s.yurchenko@ucl.ac.uk}

\author{Jonathan Tennyson}
\affiliation{Department of Physics and Astronomy, University College London, Gower Street, WC1E 6BT London, United Kingdom}

\title{The Numerical Equivalence of Diabatic and Adiabatic Representations in Diatomic Molecules}

\date{\today}

\allowdisplaybreaks

\begin{document}

\begin{abstract}

The (stationary) Schr\"{o}dinger equation for atomistic systems is solved using the adiabatic potential energy curves (PECs) and the associated adiabatic approximation. 
In cases when interactions between electronic states become important, the associated  non-adiabatic effects are taken into account via the derivative couplings (DDRs), also known as non-adiabatic couplings (NACs). For diatomic molecules, the corresponding PECs in the adiabatic representation are characterized by avoided crossings. The alternative to the adiabatic approach is the diabatic representation, obtained via a unitary transformation of the adiabatic states by minimizing the DDRs. For diatomics, the diabatic representation has zero DDR and non-diagonal diabatic couplings (DCs) ensue. The two representations are fully equivalent and so should be the rovibronic energies and wavefunctions which result from the solution of the corresponding Schr\"{o}dinger equations.

We demonstrate (for the first time), the numerical equivalence between the adiabatic and diabatic rovibronic calculations of diatomic molecules, using the \textit{ab initio} curves of yttrium oxide (YO)  and carbon monohydride  (CH)  as examples of two-state systems, where YO is characterized by a strong NAC, while CH has a strong diabatic coupling. Rovibronic energies and wavefunctions are computed using a new diabatic module implemented in variational rovibronic code \textsc{Duo}. We show that it is important to include both the Diagonal Born-Oppenheimer Correction (DBOC) and non-diagonal DDRs. We also show that convergence of the vibronic energy calculations can strongly depend on the representation of nuclear motion used and
that no one representation is best in all cases. 

\end{abstract}

\textbf{keywords:} \textit{diabatisation, non-adiabatic coupling, yttrium oxide, carbon monohydride, numerical equivalence}

\maketitle

\section{Introduction} \label{sec:intro}

Non-adiabatic effects within the electronic structure of molecules are important for many physical and chemical processes \citep{18ScStxx.diabat,07LeMaxx.diabat,22WhJiWa.diabat, 04JaXhNa.diabat,11MaKrxx.diabat,96Yarkony.diabat,17ShFaPe.diabat} such as when a chemical reaction alter electronic structure, affecting nuclear dynamics. Non-adiabatic processes are also important in astronomy and atmospheric chemistry, where collisions of free radicals and open shell molecules with spatially degenerate electronic states are often seen \citep{18KaBeVa.diabat, 16KaVaGr.diabat, 14QiJaMi.diabat, 17JaQiMi.diabat, 17DeKaVo.diabat}. Modelling electronically non-adiabatic processes has also been effective in explaining the bonding in dications such as BF$^{2+}$ \cite{95KoWrBu} and strongly ionic molecules, such as LiF \citep{81WeMexx.diabat} and NaCl \citep{22SiSoGu}, whose $^1\Sigma^+$ states show non-adiabatic behaviour.

\begin{figure*}[htbp!]
    \centering
    \includegraphics[width=0.45\textwidth]{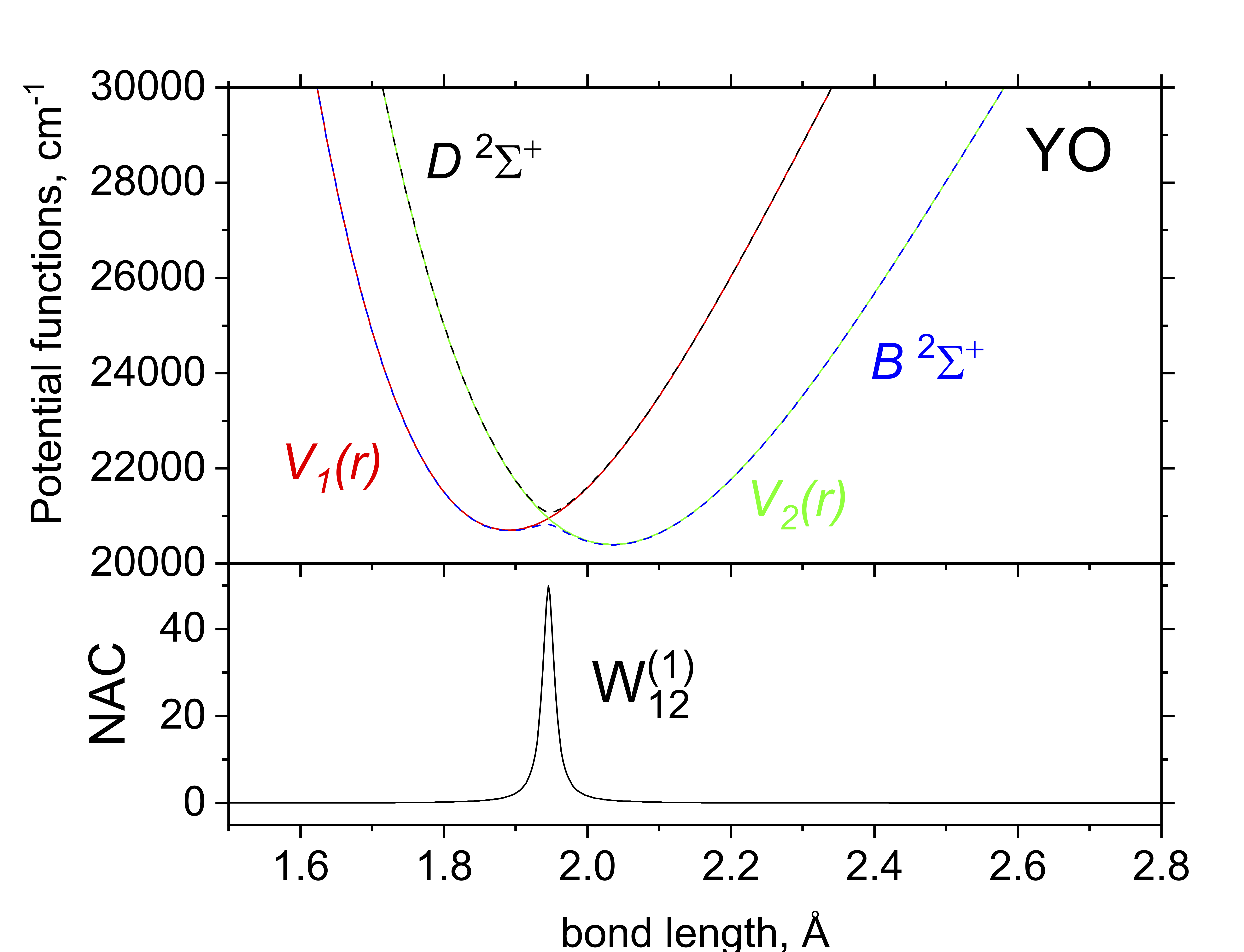}
    \includegraphics[width=0.45\textwidth]{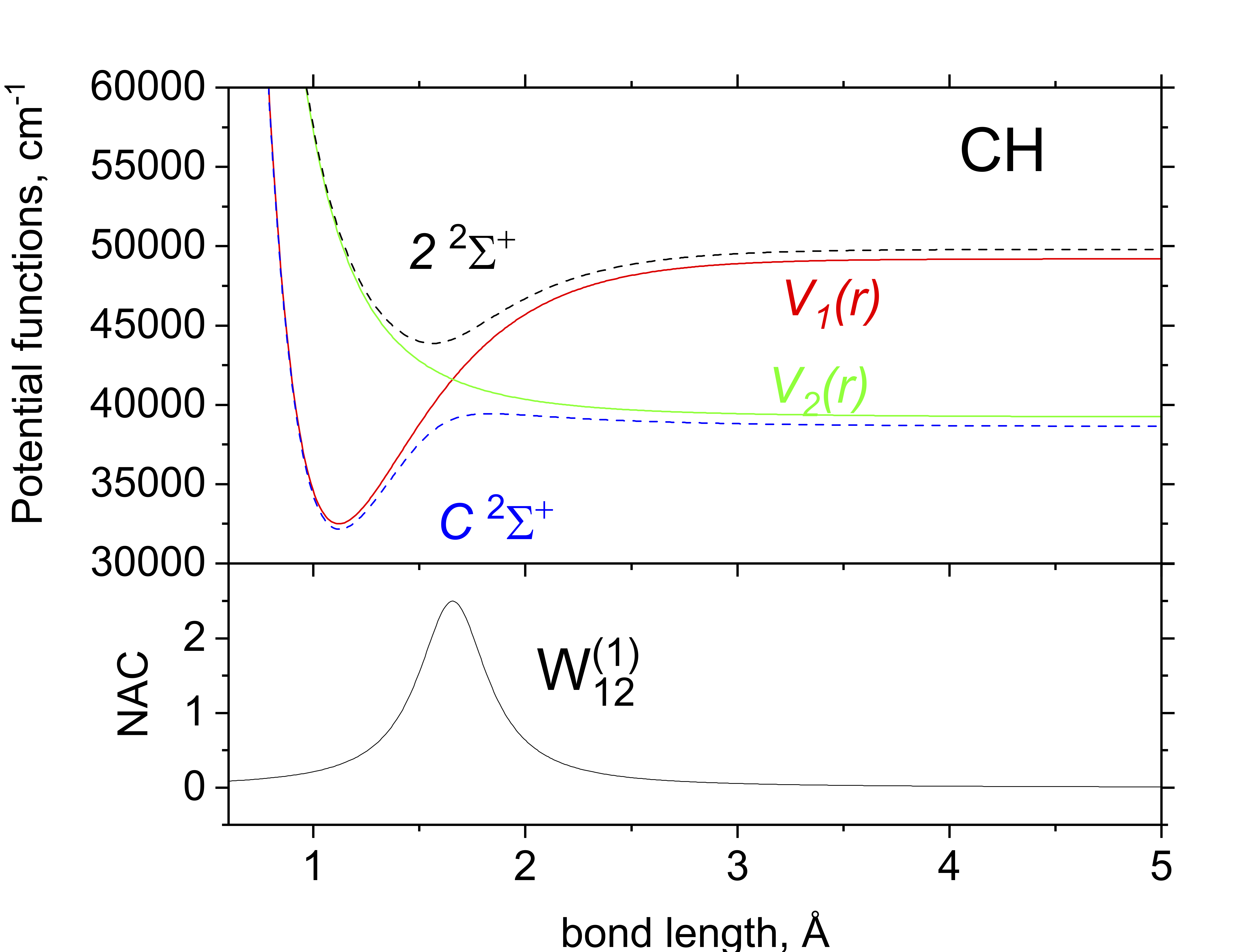}
    \caption{Illustration of the [\DS, \BS]  and [C~$^2\Sigma^+$, 2~$^2\Sigma^+$] avoided crossing systems ([black, blue] lines) for the YO and CH diatomics, respectively, which we use to perform tests on the adiabatic and diabatic equivalence. The top panels show the adiabats (solid lines) and diabats (dashed lines). The bottom panels show the corresponding NAC and DC (in the units of \AA\ and \cm) of the transformations.} 
    \label{f:YO:CH:dia}
\end{figure*}

Both the adiabatic and Born-Oppenheimer (BO) approximations assume nuclear dynamics to evolve on single electronic potential energy surfaces (PESs) \cite{18KaBeVa.diabat}, where no kinetic energy coupling (DDR) to neighbouring electronic states occurs and is generally good for predicting near-equilibrium properties for many molecules \citep{96Yarkony.diabat}. Whilst related, the adiabatic approximation differs from the BO approximation by the addition of the well-known diagonal BO correction (DBOC), introducing mass-dependence into the PECs within the adiabatic representation. The adiabatic approximation then breaks down when electronic states of same symmetry near spatial degeneracy and exhibit an avoided crossing.  \citet{20NeWi} formalised this as a non-crossing rule for diatomics, showing that potential energy curves (PECs) cannot cross and appear to `repel' upon approach (see Fig.~\ref{f:YO:CH:dia} for example). Relaxation of the BO and adiabatic approximation is then required to fully encounter the electronically non-adiabatic effects because of the inherent coupling between electronic and nuclear degrees of freedom (DoF) for both the diagonal and non-diagonal terms.



The so called derivative couplings (DDRs) or  non-adiabatic couplings (NACs) between states that exhibit avoided crossings arise through the nuclear kinetic energy operator acting on the electronic wavefunctions when the BO approximation is relaxed and corresponds to derivatives in terms of the nuclear coordinate. Computation of DDRs and PESs around the avoided crossing geometry are a major source of computational expense within both quantum chemistry and nuclear motion calculations because of the cusp-like behaviour of the PESs and the singular nature of the DDRs at the geometry of spatial degeneracy \citep{82MeTrxx.diabat,04JaKeMe.diabat,18KaBeVa.diabat,21ShVaZo.diabat}. It is therefore the main focus of many works to explore the 
    property-based 
diabatisation methods \citep{18KaBeVa.diabat,22BrYuKi,15ZhYaxx.diabat,14HoXuMa.diabat} that  transform to a diabatic  representation, where DDRs vanish or are reduced and PESs become smooth. For diatomics the smoothness condition of their PECs uniquely defines the unitary transformation to the diabatic representation where NACs (first-order non-diagonal DDR) vanish, PECs are allowed to cross and consequently the molecular properties are smooth, at the cost of introducing off-diagonal diabatic potential couplings (DC). This smoothness is then favourable for nuclear motion  calculations since no quantities within the molecular model are singular/cusped making their integration and fitting of analytical forms much simpler. The other method of diabatisation, known as point-diabatisation \citep{81WeMexx.diabat, 93RuAtxx.diabat, 97AtRuxx.diabat,01NaTrxx.diabat, 02NaTrxx.diabat, 03NaTrxx.diabat, 13XuYaTr.diabat, 08SuYeCa.diabat, 14HoXuMa.diabat,18VaPaTr.diabat}, is direct and requires the NAC being obtained \ai\ such as through the DDR-procedure \citep{Molpro} where each point can be diabatised without knowledge of the previous one, unlike property-based methods.

\citet{82MeTrxx.ai} showed that a strictly diabatic electronic basis, in which all derivative
coupling vanishes, can be defined for a diatomic system. The conditions required to make the first-order NAC vanish are straight forward,  however a true diabatic electronic basis only exists when one can remove the second-order (diagonal) derivative coupling simultaneously, which is  only possible when considering an isolated  state system, allowing one to ignore coupling to other adiabatic states. The adiabatic to diabatic transformation (AtDT) for the $N$-nuclear-coordinate case up to coupled 4-state systems has been investigated thoroughly by Baer and co-authors since the late 1980s \citep{89Baer.diabat, 00Baer.diabat, 00BaAlxx.diabat,02BaerMichael.diabat, 06Baer}. These works develop the so called line-integral approach in solution to the matrix differential equation that arises when solving for the AtDT which completely reduces the NAC matrix. Their results, albeit from a different angle to this study, are consistent with the results we present.

Despite diabatisation being used routinely to treat the avoided crossings of molecular PESs, there have been very few studies examining the numerical equivalence of adiabatic and diabatic states. This would not only be of value to those who want to benchmark their own nuclear motion codes, but to better understand the roles of each term in the diabatic and adiabatic Hamiltonian. Equivalence refers to the principle that the two representations should yield identical observables, such as energy eigenvalues. 

The solution of the nuclear motion Schr\"{o}dinger equation should not depend on whether the adiabatic or diabatic representations of the electronic states are used \citep{06Baer}. In practice with numerical applications, observables should converge to the same values with increasing accuracy of calculation, e.g by using increasingly larger basis sizes. Equivalency is often assumed but rarely shown. Convergence between the adiabatic and diabatic states have been investigated in only a handful of papers. \citet{75ZiGe} performed numerical convergence tests on adiabatic and diabatic states of the transition probability amplitudes in collisions of collinear atom–diatom systems, where the 
convergence to equivalence was demonstrated and it was  shown that  convergence was markedly different with the diabatic representation converging significantly faster.  \citet{19ShGuZh} evaluated numerical convergence rates of adiabatic and diabatic energy eigenvalues and eigenfunctions using a sinc-DVR method; equivalency was demonstrated, but this required using a complete adiabatic model and a conical intersection at high energy. 
The magnitude of the DDR corrections within the adiabatic representation has been studied before such as in the series of papers by Wolniewicz, Dressler and co-workers \citep{77WoDrxx.H2, 79DrGaQu.H2, 86DrWoxx.H2, 90QuDrWo.H2, 92WoDrxx.H2, 94WoDrxx.H2, 94YuDrxx.H2} where excited electronic states of molecular hydrogen and their coupling were studied in detail. The earliest of these studies used the adiabatic approximation but through the series non-adiabatic couplings were introduced and improved for increasing number of excited states and were shown to be essential to produce an accurate spectroscopy (i.e. accurate rovibronic energies and transitions) of the system, confirmed by comparison to experiment. In the later studies the diabatic representation was also shown to provide an accurate description of the nuclear dynamics of H$_2$, but comparisons between the adiabatic and diabatic representations were not shown. Additionally, DDR corrections were studied with respect to the computed rovibrational energies of H$_2^+$ and D$_2^+$ by \citet{08JaKuxx.diabat} and later by \citet{22Jaquet.H3+} on the H$_2^+$, H$_3^+$, and H$_2$ systems. It is therefore expected that DDR contributions are critical for the accurate determination of the energies of small hydrogen bearing  molecules. 


Non-adiabatic interactions are also important for scattering calculations which often assume the equivalence between the adiabatic and diabatic representations \citep{75Baer.diabat}. For example, \citet{jt574} provide a partial diabatic representation for the electronic structure of N$_2$ which was used within multichannel quantum defect theory calculations for the dissociative recombination of N$_2^+$ \citep{jt591} where \ai\ cross sections were generated. It was shown by \citet{15VoYaYa.diabat} that for multichannel coulomb scattering calculations for the mutual neutralisation reaction H$^+ +\;$H$^-\rightarrow\;$H$_2^*\rightarrow\;$H$(1)\;+\;$H$(n)$ that an adiabatic and diabatic reformulation produced not only equivalent results, but almost identical cross-sections as generated from various other methods. Furthermore, the influence of the second derivative coupling term was shown to be important for producing accurate cross-sections, an interesting result which showcases the need for accurate representation of non-adiabatic dynamics.


This study aims to show numerical equivalence of the adiabatic and diabatic representation in nuclear motion calculations of rovibronic energies and spectral properties for two selected diatomic systems, represented by two coupled electronic states:  yttrium oxide (YO) and carbon mono-hydride (CH) molecules illustrated in Fig.~\ref{f:YO:CH:dia}. 
YO shows avoided crossings between the \BS, \DS\ and \AS, \CS\ states as described by \citet{23YuBrTe}. YO has broad scientific interest, having been observed in stellar spectra \cite{07GoBaxx.YO,09KaScTy.YO,82Murty.YO,83Murty.YO}, found use in solar furnaces \cite{05BaCaGr.YO,07BaCaGr.YO} and magneto-optical traps \cite{15YeHuCo.YO,18CoDiWu.YO}. Yttrium oxide is a complex system showing many low-lying electronic states: accurate descriptions of its avoided crossings will be valuable to works in several fields. 
CH is one of the most studied free radicals \cite{jt868} because it occurs in such a wide variety of environments: it has been observed in flames \citep{92WiGrSe,16VeWaLi}, solar and \citep{78Lambert.C2,89MeGrSa,91GrLaSa}, 
stellar spectra  \citep{84RiCaHa,86LaGuEr,14MaPlVa.CH}, spectra of comets \citep{94WoLuWa}, ISM  \citep{37SwRo,85JuMe,93SpCr,84Lien}, molecular clouds  \citep{87StLuGe}. 
 
diaba

As part of the study we also report our implementation of the full diabatic/adiabatic treatments in our code \Duo\ \cite{Duo}, a rovibronic solver of general coupled diatomic Schr\"{o}dinger equations, which is used in the analyses.  \Duo\ is a general, open access  Fortran 2003 code\footnote{\href{https://github.com/Exomol/Duo}{https://github.com/Exomol/Duo}}.



\section{Description of the diabatisation of a two-electronic-state system}
\label{sec:Theory}

Consider a coupled two-electronic-state  system of nuclear (pure vibrational) Schr\"{o}dinger equations for a diatomic molecule in the adiabatic representation,  with the non-adiabatic effects between these two states fully accounted for,  as given by (ignoring spin and rotation angular momenta)
$$
\hat{\bf H}^{\rm (a)} {\bf \varphi}_v(r) = E_v  {\bf \varphi}_v(r), 
$$
where $r$ is the distance between the two nuclei and the Born-Huang $2\times 2$ Hamiltonian operator is (see, e.g. \citet{83Romelt.ai,18VaPaTr.diabat,19YaXiZh})
\begin{equation}
\label{e:H:adia}
  \hat{\bf H}^{\rm (a)} =   -\frac{\hbar^2}{2\mu}\begin{pmatrix}
           \frac{d^2}{dr^2}-K & \left[ \frac{d}{dr}W_{12}^{(1)}  + W_{12}^{(1)} \frac{d}{dr} \right] \\ -\left[ \frac{d}{dr}W_{12}^{(1)} + W_{12}^{(1)} \frac{d}{dr} \right] &  \frac{d^2}{dr^2}-K
          \end{pmatrix} +
          \begin{pmatrix}
            V^{\rm a}_1 & 0 \\ 0 & V^{\rm a}_2
          \end{pmatrix}.
\end{equation}
Here  $\mu = m_1 m_2/(m_1 + m_2)$ is the reduced mass, $V_1^{(a)}(r)$ and $V_2^{(a)}(r)$ are the adiabatic  potential energy functions, 
$W_{12}^{(1)}(r)$ is the first order DDR, or non-diagonal NAC, given by 
\begin{equation}
\label{eq:NAC_1DDR}
    W_{12}^{(1)} = \bra{\psi_{1}^{\rm a}}\frac{d}{dr}\ket{\psi_{2}^{\rm a}}=-\bra{\psi_{2}^{\rm a}}\frac{d}{dr}\ket{\psi_{1}^{\rm a}}
\end{equation}
where $\psi_{1}^{\rm a}$ and $\psi_{2}^{\rm a}$ are the adiabatic electronic wavefunctions, and $K(r)$ is the diagonal DDR component given by 
\begin{equation}
\label{e:K}
    K = \derivBraket{\psi^a_1}{\psi^a_1} = \derivBraket{\psi^a_2}{\psi^a_2}.
\end{equation}
Furthermore, $\frac{\hbar^2}{2\mu} K(r)$ is the well-known DBOC \citep{86HaYaSc.method}. 

The derivative coupling $K(r)$  is related to the second DDR $W_{12}^{(2)}$ through the following relations \citep{86LeYaxx.diabat,87SaYaxx.diabat}, in the $g$-, $h$-, and $k$- notations,
\begin{align}
\label{eq:ghk_notation}
g(\alpha,\beta)&=\braket{\psi^a_\alpha|\frac{d}{dr}|\psi^a_\beta} \equiv W^{(1)}_{\alpha,\beta} \\
k(\alpha,\beta)&=\derivBraket{\psi^a_\alpha}{\psi^a_\beta} \equiv K \\
\label{eq:ghk_notation2}
h(\alpha,\beta)&=\braket{\psi^a_\alpha|\frac{d^2}{dr^2}|\psi^a_\beta}=\frac{dg}{dr}-k\equiv\frac{dW^{(1)}_{\alpha,\beta}}{dr} - K = W^{(2)}.
\end{align}
In conjunction with Eqs.~(\ref{eq:ghk_notation}--\ref{eq:ghk_notation2}) and the results by \citet{06Baer,20MaZrBe.diabat,69Smith.diabat}, a simple and powerful expression for the matrix element of the diagonal DDR term $K$  for the coupled two-electronic state problem is obtained:
\begin{equation}
\label{e:w(1)=w(2)^2}
    K = \left( W_{12}^{(1)}\right)^2 = \frac{dW^{(1)}_{12}}{dr} - W^{(2)}.
\end{equation}
A diabatic representation of a two-state system can be introduced via a unitary transformation ${\bf U}(r)$ of the adiabatic electronic wavefunction vector $\boldsymbol{\psi}^{\rm a} = (\psi_1^{\rm a},\psi_2^{\rm a})^T$ in which the 1\ts{st} order DDR vanishes and PECs and other molecular properties become smooth  at the cost of introducing an off-diagonal potential energy coupling, termed a diabatic coupling (DC), between the non-adiabatically interacting electronic states \citep{81Delos.diabat, 82MeTrxx.diabat, 04JaKeMe.diabat}. The unitary $2\times 2$ matrix   ${\bf U}(r)$  is given by 
\begin{gather}
{\bf U}(\beta(r))= \begin{bmatrix} \cos(\beta(r)) & -\sin(\beta(r)) \\ \sin(\beta(r)) & \cos(\beta(r)) \end{bmatrix},
\label{eq:U(beta)}
\end{gather}
where the mixing angle $\beta(r)$ is obtained by integrating  NAC as follows \citep{99SiHaWe.diabat, 15AnBaxx.diabat, 17BaAnxx.diabat,18KaBeVa.diabat}
\begin{equation}
\label{eq:beta(r)}
\beta(r)= \beta(r_0)+\int^{r}_{-r_0} W_{12}^{(1)}(r') dr' ,
\end{equation}
where $r_0$ is a reference geometry and is usually chosen as  such that one can define a physical condition which ensures the mixing angle to equal $\pi/4$ at the crossing point $r_{\rm c}$. It can be also shown that for the diatomic one-dimensional case the transformation to a strict diabatic basis is unique and that  $W_{12}^{(1)}$  vanishes upon the diabatisation together with $K(r)$ (see Eq.~\eqref{e:w(1)=w(2)^2}). Similar to the work by \citet{84KoDoCe.diabat} who developed a Hamiltonian for the two-coupled electronic state problem, we develop theory for the diabatic and adiabatic electronic potential energy curves for the coupled two-electronic states in question. The corresponding two-electronic-state Born-Huang Hamiltonian operator $\hat{\bf H}^{\rm d}$ then becomes
\begin{equation}
\label{e:H:dia}
    \hat{\bf H}^{\rm (d)} = \begin{pmatrix}
           -\frac{\hbar^2}{2\mu} \frac{d^2}{dr^2} & 0 \\ 
          0 &  -\frac{\hbar^2}{2\mu}  \frac{d^2}{dr^2}
          \end{pmatrix} +
          \begin{pmatrix}
            V^{\rm d}_1 & V_{12}^{\rm d} \\ V_{12}^{\rm d} & V^{\rm d}_2
          \end{pmatrix},
\end{equation}
where the diabatic potential energy functions $V^{\rm d}_1(r)$ and $V^{\rm d}_2(r)$ and  the DC function $V^{\rm d}_{12}(r)$ are  given by 
\begin{equation}
{\bf V}^{\rm d}(r) = {\bf U^\dagger}{\bf V}^{\rm a}(r){\bf U} 
= \begin{pmatrix}
\centering
V^{\rm d}_1(r) & V^{\rm d}_{12}(r)\\
V^{\rm d}_{12}(r) & V^{\rm d}_2(r)
\end{pmatrix} 
= \begin{bmatrix} V^{\rm a}_1 \cos^2\beta + V^{\rm a}_2 \sin^2\beta    & \frac{1}{2}  (V^{\rm a}_{2}-V^{\rm a}_{1})  \sin(2\beta) \\  \frac{1}{2} (V^{\rm a}_{2}-V^{\rm a}_{1}) \sin(2\beta) &  V^{\rm a}_1 \sin^2\beta + V^{\rm a}_2 \cos^2\beta
\label{eq:dia_V}
\end{bmatrix}.
\end{equation}

The goal of this work is to demonstrate the equivalency of the adiabatic and diabatic representations when solving the nuclear motion diatomic (eigenvalue) problem. To this end we aim to construct, solve and compare  the eigensolutions of model diatomic systems in the adiabatic and diabatic representations.  

If the adiabatic representation of an isolated two-electronic state diatomic is fully defined by the three functions $V_1^{\rm a}(r)$, $V_2^{\rm a}(r)$ and $W_{12}^{(1)}(r)$ in Eq.~\eqref{e:H:adia},  in turn, the diabatic representation is fully defined by the three functions $V_1^{\rm d}(r)$, $V_2^{\rm d}(r)$ and $V_{12}^{\rm d}(r)$ in Eq~\eqref{e:H:dia}. In fact, both transformations can be fully described by a combination of any three functions from the set $V_1^{\rm a}(r)$, $V_2^{\rm a}(r)$, $W_{12}^{(1)}(r)$, $V_1^{\rm d}(r)$, $V_2^{\rm d}(r)$ and $V_{12}^{\rm d}(r)$. For this study we choose $V_1^{\rm d}(r)$, $V_2^{\rm d}(r)$  and $W_{12}^{(1)}(r)$. The diabatic PECs $V_1^{\rm d}(r)$, $V_2^{\rm d}(r)$  are expected to have smooth shapes by construction and are easy to parameterize,  which explains our choice, while $W_{12}^{(1)}(r)$  has also a rather simple, easy-to-parameterize cusp-like shape \citep{81WeMexx.diabat, 82MeTrxx.diabat,04JaKeMe.diabat, 18KaBeVa.diabat,21ShVaZo.diabat} as will be shown below. The other three functions are constructed from $V_1^{\rm d}(r)$, $V_2^{\rm d}(r)$  and $W_{12}^{(1)}(r)$ as follows.

We first define  $\beta(r)$ via  Eq.~\eqref{eq:beta(r)}. By applying the inverse  transformation ${\bf U}^{\dagger}$ to the  potential matrix ${\bf V}^{\rm d}(r)$ in Eq.(\ref{eq:dia_V}), we arrive at the following condition for the off-diagonal element of the adiabatic potential matrix
\begin{equation}
    \sin{\beta(r)}\cos{\beta(r)}\left(V^{\rm d}_1-V^{\rm d}_2\right)+\left( \cos{\beta(r)}^2- \sin{\beta(r)}^2 \right)V^{\rm d}_{12} = 0,
\end{equation}
which is required to be zero since ${\bf V}^{\rm a}(r) = {\bf U}{\bf V}^{\rm d}(r){\bf U^{\dagger}}$ in Eq.~\eqref{e:H:adia} is diagonal by definition. Hence, we can rearrange it for the DC to get
\begin{equation}
    V^{\rm d}_{12}=\frac{1}{2}\tan\left(2\beta(r)\right)\left(V^{\rm d}_2-V_1^{\rm d}\right).
    \label{eq:DC}
\end{equation}

The adiabatic functions $V_1^{\rm a}(r)$ are $V_2^{\rm a}(r)$ can be then constructed as eigenvalues of the diabatic potential energy matrix (second term in Eq.~\eqref{e:H:dia}):
\begin{eqnarray}
\label{e:V1:ad}
  V_{1}^{\rm a}(r) &=& \frac{V_1^{\rm d}(r)+V_2^{\rm d}(r)}{2}-\frac{1}{2}\sqrt{[V_1^{\rm d}(r)-V_2^{\rm d}(r)]^2+4 \, V_{12}^2(r)}, \\
\label{e:V2:ad}
V_{2}^{\rm a}(r) &=& \frac{V_1^{\rm d}(r)+V_2^{\rm d}(r)}{2}+\frac{1}{2}\sqrt{[V_1^{\rm d}(r)-V_2^{\rm d}(r)]^2+4 \, W_{12}^2(r)},
\end{eqnarray}
or, equivalently, via the inverse unitary transformation ${\bf U}$:
\begin{equation}
{\bf V}^{\rm a}(r) = {\bf U}{\bf V}^{\rm d}(r){\bf U}^\dagger 
= \begin{pmatrix}
\centering
V^{\rm a}_1(r) & 0\\
0 & V^{\rm a}_2(r)
\end{pmatrix} 
= \begin{bmatrix} V^{\rm d}_1 \cos^2\beta + V^{\rm d}_2 \sin^2\beta    & 0 \\  0 &  V^{\rm d}_1 \sin^2\beta + V^{\rm d}_2 \cos^2\beta
\label{eq:adia_V}
\end{bmatrix}.
\end{equation}

\section{Spectroscopic Models} \label{sec:spectroscopic models}

As an illustration, two model two-state electronic systems are used, YO and CH with their diabatic and adiabatic curves shown in Fig.~\ref{f:YO:CH:dia} and introduced in detail in the following. 

\subsection{YO spectroscopic model}

As an example of a two-state system with narrow, coupled bound electronic curves, we choose the \ai\ PEC curves of the \BS\ and \DS\ states of yttrium oxide from \citet{19SmSoYu} with the NAC  from \citet{23YuBrTe}. 

We use a  Morse oscillator  function as a simple model for the diabatic \BS\ and \DS\  PECs of  YO as given by  \begin{equation}\label{e:Morse}
V(r)=T_{\rm e}\;\;+\;\;(A_{\rm e} - T_{\rm
e})\, \left[1-\exp\left(-b\, (r-r_{\rm e})
\right)\right]^2,
\end{equation}
where $A_{\rm e} $ is a dissociation asymptote,  $A_{\rm e} - V(r_{\rm e})$ is the dissociation energy and  $r_{\rm e}$ is an equilibrium distance of the PEC. The NAC of YO   can be efficiently described by a  Lorentzian function:
\begin{equation}
W_{12}^{(1)}(r) =  \frac{1}{2}\frac{\gamma}{\gamma^2+(r-r_{\rm c})^2},
\label{e:lorentzian}
\end{equation}
where $\gamma$ is the corresponding half-width-at-half-maximum (HWHM), while $r_{\rm c}$ is its center, corresponding to the crossing point of diabatic curves. 
These PECs and NAC are illustrated in Fig.~\ref{f:YO:dia:adia}. The  parameters defining these curves are listed in Table~\ref{t:YO:params}, which were obtained by fitting to the corresponding \ai\ data. 

For the Lorenztian as a NAC, Eq.~\eqref{eq:beta(r)} is easily integrable to give the transformation angle $\beta(r)$:
\begin{equation}
\beta(r) =\frac{\pi}{4}\pm \frac{1}{2}\arctan(\frac{r-r_{\rm c}}{\gamma}),
\end{equation}
where  $r_{\rm c}$ is obtained as the  crossing point between the  PECs, and the $\pm$ sign refers to the path integral when $r<r_{\rm c}$ and $r_{\rm c}<r$ respectively.

The adiabatic curves obtained using Eqs.~(\ref{e:V1:ad},\ref{e:V2:ad}) and the DC curve obtained using Eq.~\eqref{eq:DC} are shown in Fig.~\ref{f:YO:dia:adia}.  The value of the crossing point $r_{\rm c}$  is obtained as numerical solution of $V_{1}^{\rm d} = V_{2}^{\rm d}$ and is listed in Table~\ref{t:YO:params}. 
The  derivative coupling $K$ in the diagonal matrix element of the  adiabatic  kinetic energy operator in Eq.~\eqref{e:H:adia} is simply defined by $K = \left({W_{12}^{(1)}}\right)^2$ according to Eq.~\eqref{e:w(1)=w(2)^2}. All the corresponding curves are programmed in \Duo\ analytically and are  provided on a grid of 1000 equidistant bond lengths as part of the supplementary material.
\begin{table}[htbp!]
\footnotesize
    \centering
    \caption{The molecular parameters defining the YO spectroscopic model}
    \label{t:YO:params}
    \begin{tabular}{lccc}
        \hline
        Parameter &  $V_1^{\rm d}$ & $V_2^{\rm d}$ & $W_{12}^{(1)}$ \\ 
        \hline\hline 
$T_{\rm e}$, \cm &  20700.0  & 20400.0 \\ 
$r_{\rm e}$, \AA &  1.89     &  2.035   \\
$b$, \AA$^{-1}$  &  1.5      &  1.26    \\
$A_{\rm e}$, \cm &  59220.0  &  59220.0 \\
$\gamma$ , \cm  &  & &   0.01  \\
$r_{\rm c}$, \AA   &  & &        1.945843834  \\
        \hline
    \end{tabular}
\end{table}

\subsection{CH spectroscopic model}

The spectroscopic model for CH, with curves illustrated in Fig.~\ref{f:YO:CH:dia} (right panel), is constructed to mimic the \ai\ curves of \Cstate\ and \Zstate\ by \citet{87vanDishoeck}. The \Cstate\ state has a bound shape with a well of about 16700~\cm\ (2.0705~eV), which we model using a Morse oscillator function in Eq.~\eqref{e:Morse}. The \Zstate\ state is repulsive, with the dissociation energy lower than that of  \Cstate\ by about 10000~\cm. We chose to model the \Zstate\ PEC using the following form:
\begin{equation}
    \label{e:1/r^4}
    V(r) = D_{\rm e} + C_4/r^4. 
\end{equation}
The corresponding NAC between \Cstate\ and \Zstate\ of CH from \citet{87vanDishoeck} is modelled using a two-parameter Lorentzian function in Eq.~\eqref{e:lorentzian}. All parameters defining the CH spectroscopic model are listed in Table~\ref{t:CH:params}. As above, the value of the crossing point $r_{\rm c}$  is obtained as a numerical solution of $V_{1}^{\rm d} = V_{2}^{\rm d}$. 

\begin{table}[htbp!]
\footnotesize
    \centering
    \caption{The molecular parameters defining the CH diabatic  spectroscopic model}
    \label{t:CH:params}
    \begin{tabular}{lcccc}
        \hline
        Parameter &  $V_1^{\rm d}$ & $V_2^{\rm d}$ & \Xstate &  $W_{12}^{(1)}$ \\ 
        \hline\hline 
$T_{\rm e}$, \cm    &  32500.0  & & 0.0  \\ 
$r_{\rm e}$, \AA    &  1.12     &  & 1.12    \\
$b$, \AA$^{-1}$     &  2.5     &  & 1.968  \\
$A_{\rm e}$, \cm    &  49200.0  &  29374.0   &  39220.0 \\
$C_4$, \AA$^{-4}$   & & 18000.0  \\
$\gamma$ , \cm  &   & &  &  0.2  \\
$r_{\rm c}$, \AA    &  & &  & 1.656644935  \\
        \hline
    \end{tabular}
\end{table}

\section{Solving the rovibronic Schr\"{o}dinger equations for CH and YO}

Both the CH and YO doublet systems represent open-shell molecules. Towards a complete rovibronic solution, the pure vibrational Hamiltonian operator in Eqs.~\eqref{e:H:adia} or \eqref{e:H:dia} are extended with the rotation-spin-electronic contribution as follows (see \citet{Duo} for details of the  approach used): 
\begin{equation}
\hat{H} = \hat{H}_{\rm vib} + \frac{\hbar^2}{2\mu} \hat{R}^2,
\end{equation}
where the rotational angular momentum operator $\hat{R}$ is replaced with 
\begin{equation}
\hat{R} = \hat{J} - \hat{S} - \hat{L}. 
\end{equation}
Here $\hat{J}$, $\hat{S}$,  $\hat{L}$ are  the total, spin and electronic angular momenta, respectively. We then solve the aforementioned rovibronic Schr\"{o}dinger systems for YO and CH variationally on  the Hund's case (a) basis using the \Duo\ program \cite{Duo}, which has been extended as part of this work to include the adiabatic and diabatic effects. 
The spectroscopic models of CH and YO are provided in the form of the \Duo\ input files both in the diabatic and adiabatic representations as the supplementary material. 

\Duo\ uses the numerical sinc-DVR method \citep{82GuRoxx,84LuRixx} to solve the  Schr\"{o}dinger systems for the curves defined either on a grid or as analytic functions. For the analytic representations as above, the corresponding functions are mapped on a grid of sinc-DVR points. For a grid input, cubic splines are used. The \Duo\ kinetic energy has been extended to include the first derivative component required for implementation of the NAC, also using the sinc-DVR representation \citep{06LoSh}. The DBOC terms can be either provided as  input or generated from the NAC using Eq.~\eqref{e:K}. In order to facilitate numerically exact equivalency of the diabatic and adiabatic representations \Duo\ calculations, Eqs.~(\ref{eq:DC}, \ref{e:V1:ad} and \ref{e:V2:ad}) are provided are used for constricting $V^{\rm d}_{12}$, $V_{1}^{\rm a}(r)$ and $V_{2}^{\rm a}(r)$, respectively, from $V_{1}^{\rm d}(r)$, $V_{2}^{\rm a}(r)$ and $\beta(r)$.

\subsection{The YO solution}

\begin{figure}[htbp!]
    \centering
    \includegraphics[width=0.90\textwidth]{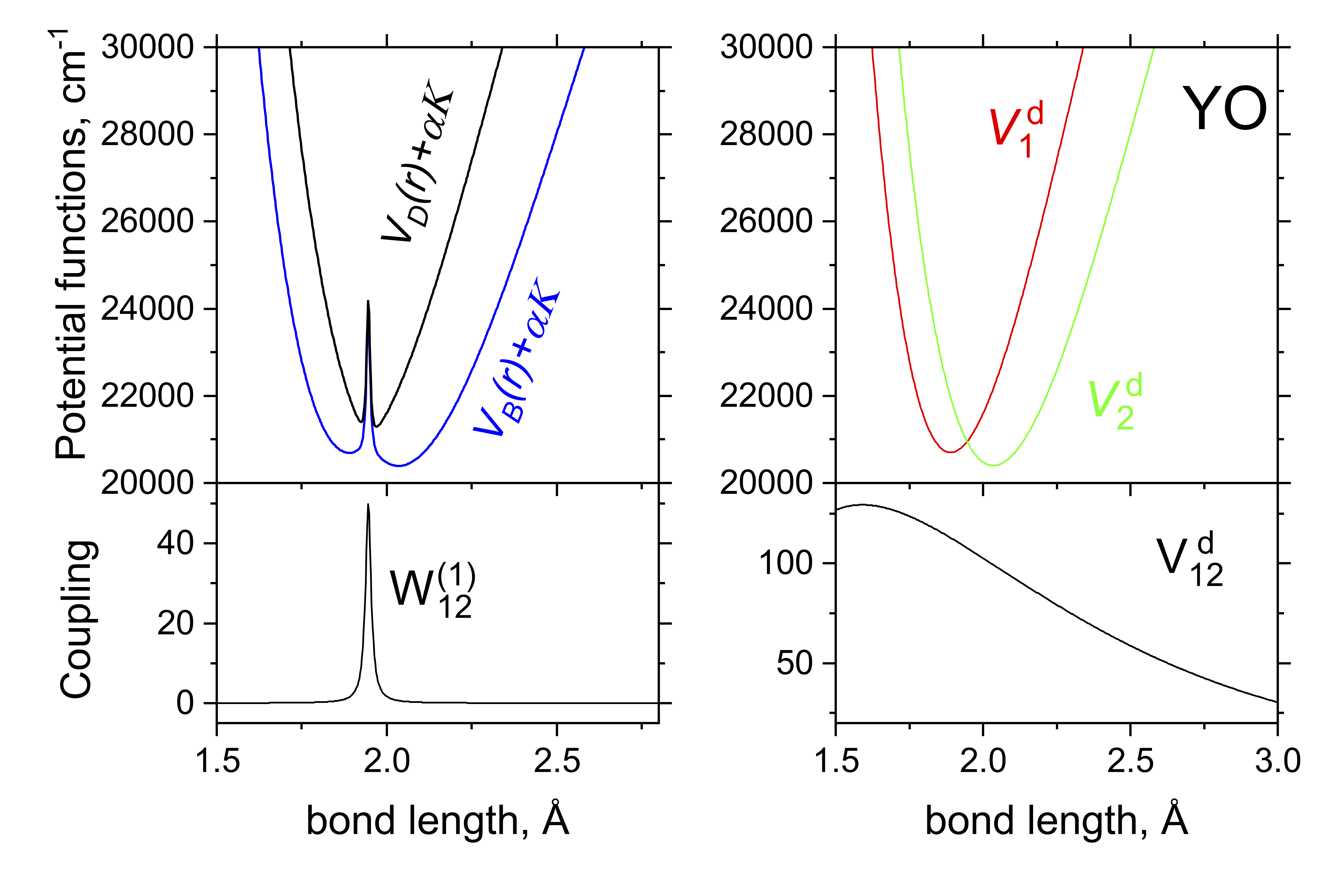}
    \caption{Full adiabatic (left) and diabatic (right) models of the \BS\ and \DS\ systems of YO. The top panels show the PECs, where the  adiabatic PECs include the diagonal DDR correction $\alpha K$ and $\alpha = h/(8 \pi^2 c \mu)$.  The bottom panels show the corresponding coupling curves, NAC (left)  and DC (right).} 
    \label{f:YO:dia:adia}
\end{figure}

We first find the vibronic ($J=0.5$) energies  of the coupled \BS\ and \DS\ systems in  the adiabatic and diabatic representations as accurately as possible in order to establish a baseline and also to demonstrate the equivalency of the two representations. Even though we know that the diabatic and adiabatic solutions should be equivalent (i.e. identical within the calculation error), this is always a subject to the convergence or other numerical limitations. For example,  \Duo\ uses a PEC-adapted vibrational basis set constructed by solving  the pure vibrational problem, which will be different depending on the representation, diabatic or adiabatic, and thus will influence the convergence. The corresponding YO model curves are shown in Fig.~\ref{f:YO:dia:adia}, where the DBOC coupling $K$ is included into the adiabatic PECs for clarity. There is a striking difference between the two models, with a large spike in the middle of the adiabatic PECs, yet we expect them to give the same eigenvalues and eigenfunctions. 

A selected set of rovibronic energy term values ($J=0.5$) computed using the two methods is listed in Table~\ref{t:YO:energies:J0.5}. The energies are indeed identical (within $2.5\times10^{-5}$~\cm), but the approximate quantum state  labels as assigned by \Duo\ are very different. \Duo\ assigns quantum labels via the largest contribution from the corresponding basis sets, which in both cases are very different and so are their state interpretations, in which case we compare states of matching energy enumerator $n$.

Having established the numerical equivalence, we can now investigate the importance of different non-adiabatic couplings for the YO model. Three approximations are considered here: (A1) in the adiabatic model, both DDR terms are switched off ($W_{12}^{(1)}= K = 0$); (A2) in the adiabatic model, the diagonal DDR is switched off ($K = 0$), but the NAC is kept in; (A3) in the diabatic model, the diabatic coupling is set to zero ($V_{12}=0$). The effects of  these approximations on the calculated energies of YO ($J=0.5$) are also shown in Table~\ref{t:YO:energies:J0.5}. 
For the adiabatic model, the omission of $K$ (A2) has the overall largest impact especially on the \BS\ term values. The omission of $V_{12}$ from the diabatic model (A3) appears to be  less damaging than other two approximations. It is clear, however, that any degradation of theory leads to large errors, unacceptable for  high-resolution applications. This is in fact the main conclusion of this work that the impact of  dropping any of non-adiabatic corrections from the model describing a system with crossing has to be always investigated. 

Out of the two representations, the adiabatic model is usually considered to be more complex to work with. Its curves have complex shapes with the model being very sensitive to the mutual consistency of the curves $V_{1}^{\rm a}$, $V_{2}^{\rm a}$ and $W_{12}^{(1)}$ around the crossing point. The disadvantage of the diabatic representation is that it does not come out as a solution of the (adiabatic) electronic structure calculations directly and needs to be constructed either through a  diabatisation approach \citep{18KaBeVa.diabat,22BrYuKi,15ZhYaxx.diabat,14HoXuMa.diabat,81WeMexx.diabat, 93RuAtxx.diabat, 97AtRuxx.diabat,01NaTrxx.diabat, 02NaTrxx.diabat, 03NaTrxx.diabat, 13XuYaTr.diabat, 08SuYeCa.diabat,18VaPaTr.diabat} or approximated.

\begin{table}[htbp!]
\footnotesize
    \centering
    \caption{The rovibronic ($J=0.5$) energy term values (\cm) of the \BS\ (B) and \DS\ (D) systems  of YO computed using the adiabatic and diabatic representations. The energies are listed relative to the lowest $J=0.5$ state.  }
    \label{t:YO:energies:J0.5}
    \begin{tabular}{rrrrcrcrrcr}
        \hline\hline 
        $n$& \multicolumn{5}{c}{Adiabatic} && \multicolumn{4}{c}{Diabatic} \\
        \cline{1-6} \cline{8-11}
& $\tilde{E}$  &  $\tilde{E}$(DDRs=0) & $\tilde{E}(K=0)$ & State   &  $v$  &  &  $\tilde{E}$  &  $\tilde{E}(V_{12}=0)$   &  State   &    $v$      \\
        \cline{1-6} \cline{8-11}
     1 &     0.000000 &       0.000000 &       0.000000 &   B   &  0   &  &       0.000000 &     0.000000 &   D   &   0   \\
     2 &   344.431810 &     347.928597 &     191.831751 &   B   &  1   &  &     344.431809 &   351.249676 &   B   &   0   \\
     3 &   561.079914 &     690.986320 &     492.221984 &   B   &  2   &  &     561.079921 &   549.732652 &   D   &   1   \\
     4 &  1009.133229 &     967.537324 &     983.098980 &   B   &  3   &  &    1009.133232 &  1002.246089 &   B   &   1   \\
     5 &  1108.354299 &    1132.062465 &    1129.463766 &   D   &  0   &  &    1108.354283 &  1095.516787 &   D   &   2   \\
     6 &  1612.539760 &    1553.296745 &    1777.897073 &   B   &  4   &  &    1612.539736 &  1637.352406 &   D   &   3   \\
     7 &  1688.323434 &    1897.761066 &    1868.635701 &   B   &  5   &  &    1688.323453 &  1647.646531 &   B   &   2   \\
     8 &  2179.350796 &    2008.167697 &    2345.749886 &   D   &  1   &  &    2179.350783 &  2175.239507 &   D   &   4   \\
     9 &  2297.569318 &    2465.488852 &    2396.923772 &   B   &  6   &  &    2297.569321 &  2287.451003 &   B   &   3   \\
    10 &  2718.929830 &    2689.784491 &    2839.568147 &   B   &  7   &  &    2718.929830 &  2709.178092 &   D   &   5   \\
    11 &  2928.147305 &    2925.374682 &    3115.611400 &   D   &  2   &  &    2928.147294 &  2921.659505 &   B   &   4   \\
    12 &  3247.771603 &    3395.227251 &    3377.138924 &   B   &  8   &  &    3247.771603 &  3239.168161 &   D   &   6   \\
    13 &  3559.124439 &    3442.432354 &    3666.238711 &   D   &  3   &  &    3559.124429 &  3550.272037 &   B   &   5   \\
    14 &  3772.447582 &    3862.695406 &    3963.866748 &   B   &  9   &  &    3772.447578 &  3765.209712 &   D   &   7   \\
    15 &  4181.801597 &    4167.979957 &    4373.535285 &   D   &  4   &  &    4181.801594 &  4173.288598 &   B   &   6   \\
    16 &  4295.897860 &    4333.054560 &    4472.298326 &   B   &  10  &  &    4295.897854 &  4287.302747 &   D   &   8   \\
    17 &  4783.958004 &    4805.617146 &    4913.118506 &   B   &  11  &  &    4783.958001 &  4790.709188 &   B   &   7   \\
    18 &  4829.238038 &    4866.961640 &    4961.045768 &   D   &  5   &  &    4829.238030 &  4805.447266 &   D   &   9   \\
    19 &  5320.626170 &    5275.859430 &    5497.071432 &   B   &  12  &  &    5320.626156 &  5319.643267 &   D   &  10   \\
    20 &  5417.844769 &    5552.275088 &    5610.459386 &   D   &  6   &  &    5417.844772 &  5402.533809 &   B   &   8   \\
   
    & \ldots & \ldots &  \ldots  & & & & \ldots  &  \ldots   \\
        \hline
    \end{tabular}
\end{table}

\subsection{Eigenfunctions and Reduced Density} 
\label{subsec:Reduced Density}

It is  instructive to compare the eigenfunctions $\varphi_{i}^{J,\tau}(r)$ of the adiabatic and diabatic solutions and different approximations. To this end we form reduced radial densities of  the eigen-state in question. The eigenfunctions $\varphi_{i}^{J,\tau}$ utilised by Duo are expanded in the basis set $\ket{n}$,
\begin{equation}
    \label{e:phi}
    \varphi_{i}^{J,\tau} = \sum_{n=1}^{N} C_{i,n}^{J,\tau} \ket{n},
\end{equation}
where $N$ is the basis size and $C_{i,n}^{J,\tau}$ are the expansion coefficients used to assign quantum numbers by largest contributions. $\ket{n}$ denotes the full basis: $\ket{n} = \ket{{\rm st}, J, \Omega, \Lambda, S, \Sigma, v}$ where `st' is the electronic state, $S$ is the electron spin angular momentum, $v$ is the vibrational quantum number, and $\Lambda$, $\Sigma$ and $\Omega$ are the projections of electron orbital, spin, and total  angular momentum along the internuclear axis respectively. The reduced radial density $\rho_{i}^{J,\tau}(r) $ is then given by 
\begin{equation}
    \label{e:rho(r)}
    \rho_{i}^{J,\tau}(r) = \sum_{v} \sum_k | C_{v,k}^{i,J,\tau} |^2 \, | \chi_v (r) |^2 ,
\end{equation}
where $\ket{k} =  \ket{st, J, \Omega, \Lambda, S, \Sigma}$ and $\chi_v (r)$ are the vibrational wavefunctions. The reduced density states are probability density functions over bond length averaged over all quantum numbers in $\ket{n}$. This is an efficient way of examining the behaviour of the wavefunctions without looking in a hyperdimensional space defined by quantum numbers $\ket{n}$.

\begin{figure*}[ht!]
    \centering
    \includegraphics[width=.48\textwidth]{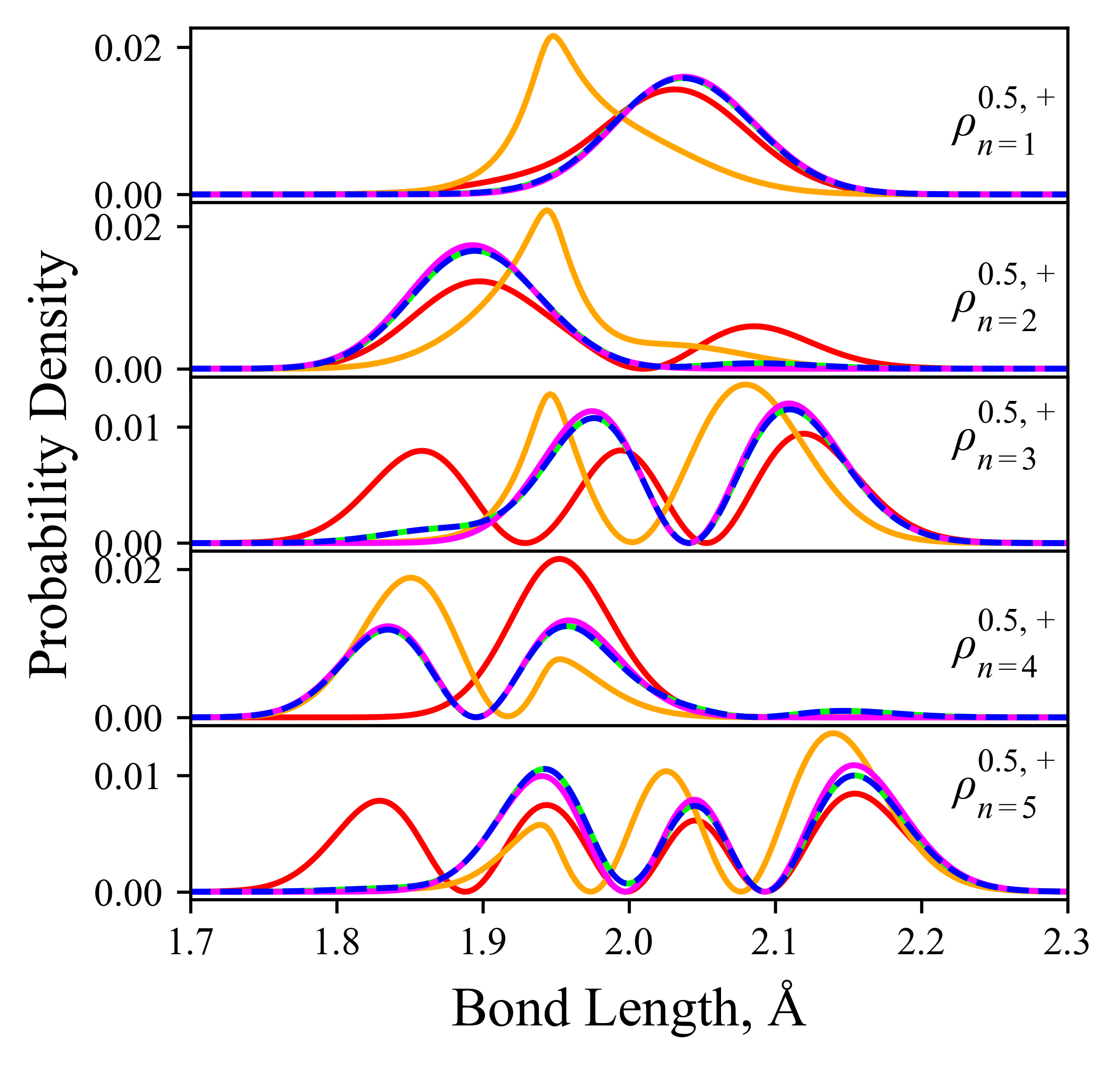}
    \caption{YO reduced density states for the lowest 5 bound levels with n being the energy enumerator given in Table \ref{t:YO:energies:J0.5}. These reduced densities are illustrated and computed using different levels of theory: diabatic representation with DC (blue dotted); diabatic model with the DC turned off (magenta, A3); adiabatic representation with both the NAC and $K$ correction included (lime green); adiabatic representation with NAC only (orange, A2); adiabatic representation with no correction (red, A1).} 
    \label{f:rho:YO}
\end{figure*}

Figure \ref{f:rho:YO}  shows selected reduced radial state densities of YO computed using different representations and approximations. As expected from our energy comparisons,  the  diabatic  and  adiabatic representations  produce identical results, whereas the reduced densities quickly deviate when omitting the NAC and/or $K$ corrections. Again, it appears that the adiabatic representation with approximations  is almost better when the DDRs are completely omitted rather than omitting only one, at least concerning the lower energy levels.


\section{Adiabatic and diabatic solutions for CH}

\begin{figure}[htbp!]
    \centering
    \includegraphics[width=0.90\textwidth]{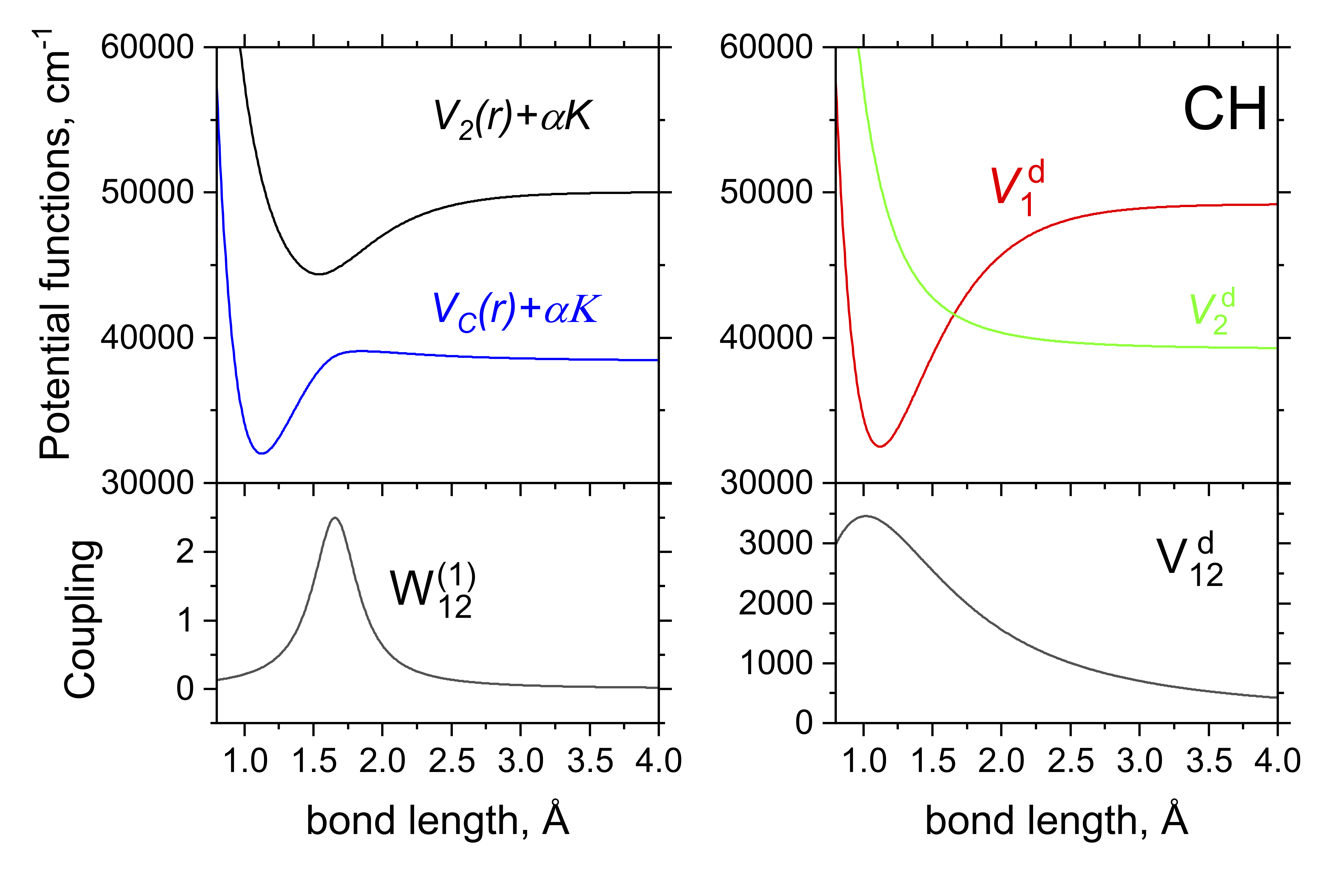}
    \caption{Full adiabatic (left) and diabatic (right) models of the \Cstate\ and \Zstate\ systems of CH. The top panels show the PECs, where the  adiabatic PECs include the diagonal DDR correction $\alpha K$, where $\alpha = h/(8 \pi^2 c \mu)$. The bottom panels show the corresponding coupling curves, NAC (left)  and DC (right).} 
    \label{f:CH:dia:adia}
\end{figure}

We now turn to a slightly different system of the \Cstate\ and \Zstate\ states of $^{12}$CH shown in Fig.~\ref{f:CH:dia:adia}. Adiabatically, these states have a large separation and a broad NAC. In contrast to YO, there is no spike-type contribution from the DBOC-term $K$ to the adiabatic PECs of CH. Diabatically, the system consists of a bound and a repulsive state with a crossing at large distance and high energy which therefore should not influence the lower rovibronic states  of \CS\ significantly. Regardless of the representation used, the region above the first dissociation channel (39220.0~\cm) is heavily  (pre-)dissociated and should contain both (pre-)dissociative and continuum states. \Duo\ is capable of finding both the bound and continuum eigensolutions. While the bound wavefunctions satisfy the standard  boundary condition to decay at large and short distances, the continuum wavefunctions can be also computed  with the sinc-DVR method used by \Duo\ and satisfy the  boundary condition of vanishing exactly at the simulation box borders (together with its first derivatives), see \citet{22PeTeYu}. 
For the analysis we separate the (quasi-)bound and the continuum states by checking the character of the wavefunctions at the `right' border $r_{\rm max}$: while the continuum states tend to oscillate at $r\to \infty$   with a non-zero density around  $r_{\rm max}$ \citep{22YuNoAz} where the bound state vanish completely. 

The resulting energy term values of the bound states are listed in Table~\ref{t:CH:energies} for all five cases, including non-adiabatic and diabatic couplings are considered as in the YO example. The full diabatic and adiabatic (bound) \Cstate\ energies  are fully equivalent within $10^{-6}$~\cm\ (here shown up to the second decimal point). However any degradation of the theory leads to drastic changes in the topology of the system and hence in the calculated rovibronic energies of the \Cstate\ state, with the error quickly deteriorating already for $v=2$. For example, removing the DC term, the diabatic solution becomes  meaningless with lots of non-physical bound states above the first dissociation channel, non-existent in the case of the full treatment. A similar effect is caused by the omission of the derivative couplings from  the adiabatic pictures with bound spurious \Zstate\ states produced by the adiabatically bound PEC \Zstate\ (see Fig.~\ref{f:CH:dia:adia}). Although the omission of the $K(r)$ term from the adiabatic solution seems harmless for the topology of the corresponding PECs, even this case leads to a spurious vibrational \Zstate\ ($v=0$) state. Therefore the conclusion is that every non-adiabatic term should be considered important, unless proven otherwise.

\begin{figure*}[ht!]
    \centering
    \includegraphics[width=.48\textwidth]{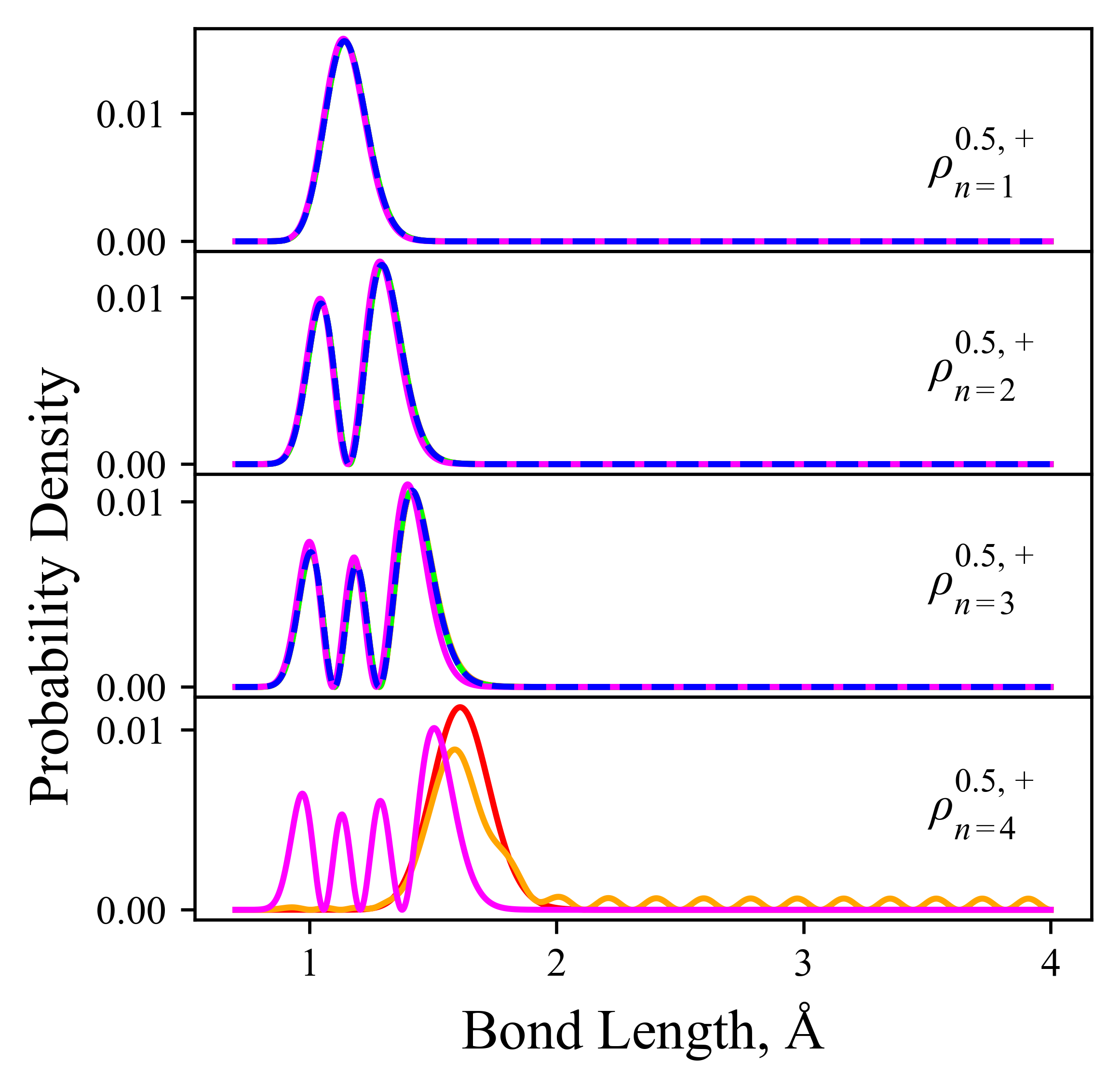}
    \caption{CH reduced density states for the lowest four bound rovibronic levels with $n$ being the energy enumerator given by the row number in Table \ref{t:CH:energies}. Different levels of theory are used to compute these reduced densities and are illustrated: diabatic representation with DC (blue dotted); diabatic model with the DC turned off (magenta, A3); adiabatic representation with both the NAC and $K(r)$ correction included (lime green); adiabatic representation with NAC only (orange, A2); adiabatic representation with no correction (red, A1).} 
    \label{f:rho:CH}
\end{figure*}

The corresponding reduced densities for some lower lying bound states of CH (\Cstate, $J=0.5$) are shown in Fig.~\ref{f:rho:CH} ($n=1,2,3$). We see that the low lying vibronic states of \CS\ are largely unaffected by the omission of the DDRs or DC since they are energetically well separated from the region of non-adiabatic interaction, in this case occuring near dissociation. However, the reduced densities of the \Zstate\ state ($n=4$) quickly diverge when removing the NAC and/or $K$ correction. The \Zstate\ is adiabatically bound and diabatically unbound, where this drastic difference is seen in the reduced densities of Fig.~\ref{f:rho:CH}, and correspond to energy levels which arise from PECs of very different character. For example, in the diabatic case where the DC is omitted, the $n=4$ state corresponds to the bound \Cstate$(J=0.5,+,v=3)$ state whereas in the adiabatic A1 and A2 cases the $n=4$ bound state corresponds to the bound \Zstate$(0.5,+,v=0)$ state. In the cases where the DDRs and DC are fully accounted for no fourth bound state exists since the couplings will push it into the quasi-bound region about the adiabatic potential hump of the \Cstate\ state. This quasi-bound nature begins to show itself in the reduced density of the adiabatic case with $K=0$ where small oscillations propagating to the right simulation border at 4~\AA\ are seen.

\begin{table}[htbp!]
\footnotesize
    \centering
    \caption{The rovibronic ($J=0.5, 1.5$ and $2.5$) bound energy term values (\cm) of the \Cstate\ (C) and \Zstate\ (2) systems  of CH computed using the adiabatic and diabatic representations.  The energies are listed relative to the lowest $J=0.5$ state.}
    \label{t:CH:energies}
    \begin{tabular}{rcccrrrcccrr}
        \hline\hline 
  $J$ &$e/f$ & \multicolumn{5}{c}{Adiabatic} && \multicolumn{4}{c}{Diabatic} \\
        \cline{1-7} \cline{9-12}
  &  &  State   &  $v$   &  $\tilde{E}$  &  $\tilde{E}$(DDRs=0) & $\tilde{E}(K=0)$ &  &   State   &    $v$  &  $\tilde{E}$  &  $\tilde{E}(V_{12}=0)$       \\
        \cline{3-7} \cline{9-12}
   0.5 &  e  &   C   &  0  &      0.00 &      0.00 &      0.00 &  &   C   &   0 &      0.00 &      0.00   \\
   0.5 &  e  &   C   &  1  &   2450.23 &   2448.12 &   2446.42 &  &   C   &   1 &   2450.23 &   2524.70   \\
   0.5 &  e  &   C   &  2  &   4617.30 &   4608.42 &   4601.48 &  &   C   &   2 &   4617.30 &   4822.76   \\
   0.5 &  e  &   2   &  0  &           &  11191.50 &  13607.15 &  &   C   &   3 &           &   6894.18   \\
   0.5 &  e  &   2   &  1  &           &  12464.33 &           &  &   C   &   4 &           &   8738.95   \\
   0.5 &  e  &   2   &  2  &           &  13549.98 &           &  &   C   &   5 &           &  10357.08   \\
   0.5 &  e  &   2   &  3  &           &  14449.75 &           &  &   C   &   6 &           &  11748.57   \\
   0.5 &  e  &   2   &  3  &           &           &           &  &   C   &   7 &           &  12913.41   \\
   0.5 &  e  &   2   &  3  &           &           &           &  &   C   &   8 &           &  13851.60   \\
   0.5 &  e  &   2   &  3  &           &           &           &  &   C   &   9 &           &  14563.15   \\
   0.5 &  f  &   2   &  2  &     27.83 &     27.82 &     27.82 &  &   C   &   6 &     27.83 &     28.08   \\
   0.5 &  f  &   2   &  3  &   2476.23 &   2474.10 &   2472.39 &  &   C   &   7 &   2476.23 &   2551.11   \\
   0.5 &  f  &   C   &  0  &   4641.14 &   4632.21 &   4625.24 &  &   C   &   8 &   4641.14 &   4847.45   \\
   0.5 &  f  &   2   &  1  &           &  11205.63 &  13620.21 &  &   C   &   9 &           &   6917.10   \\
   0.5 &  f  &   2   &  2  &           &  12478.11 &           &  &   C   &   0 &           &   8760.06   \\
   0.5 &  f  &   2   &  3  &           &  13562.86 &           &  &   C   &   1 &           &  10376.30   \\
   0.5 &  f  &   2   &  4  &           &  14461.35 &           &  &   C   &   2 &           &  11765.83   \\
   0.5 &  f  &   2   &  4  &           &           &           &  &   C   &   3 &           &  12928.61   \\
   0.5 &  f  &   2   &  4  &           &           &           &  &   C   &   4 &           &  13864.61   \\
   0.5 &  f  &   2   &  4  &           &           &           &  &   C   &   5 &           &  14573.78   \\
        \hline
    \end{tabular}
\end{table}


\subsection{Continuum solution of CH: photo-absorption spectra}

\begin{figure}[htbp!]
    \centering
    \includegraphics[width=0.90\textwidth]{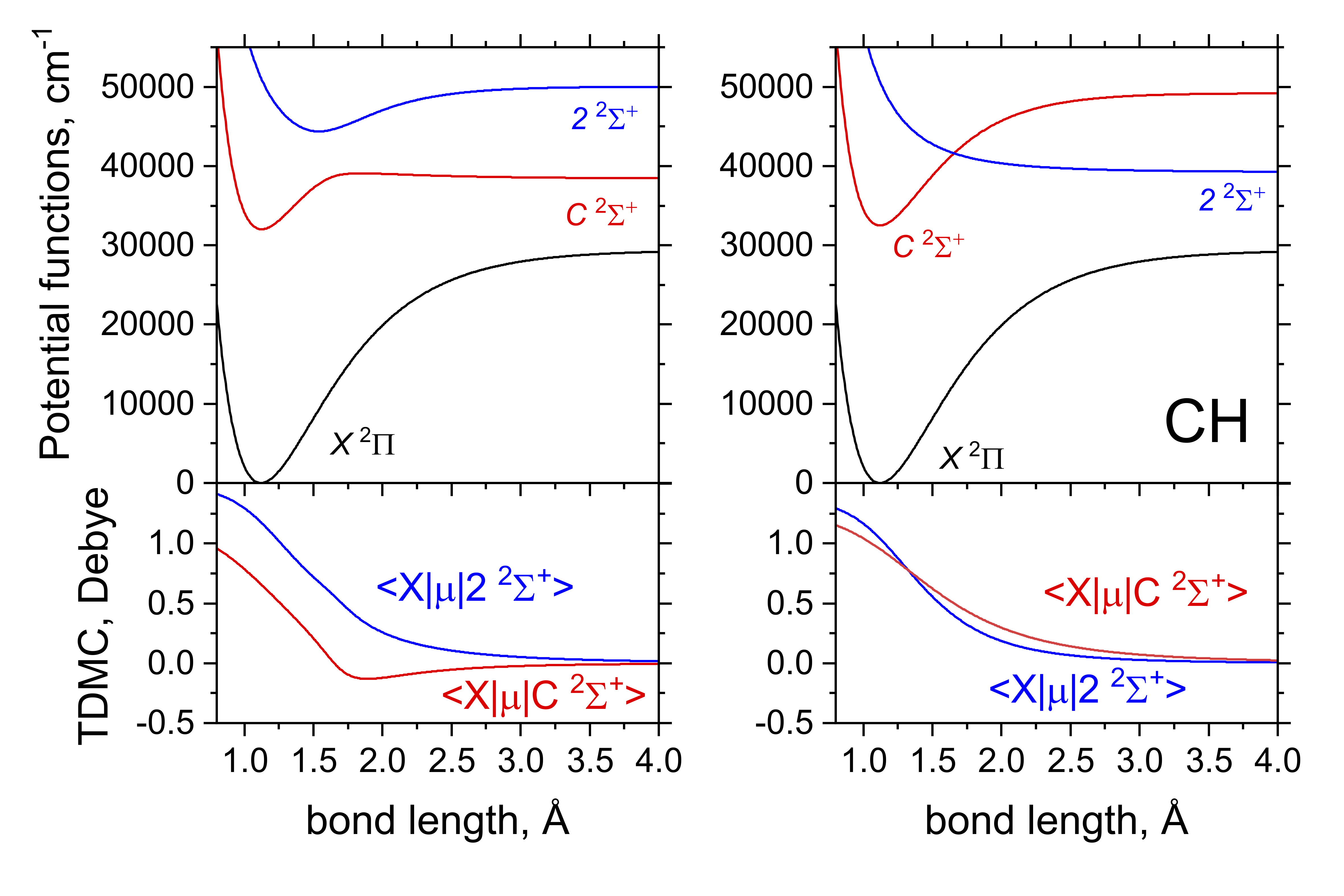}
    \caption{Adiabatic (left) and diabatic (right) models of the photo-absorption system of \Xstate\ $\to$ \Cstate/\Zstate\ of CH. The top panels show the PECs, adiabatic and diabatic, while the bottom panels show the corresponding transition dipole moment curves.} 
    \label{f:CH:X-C-2}
\end{figure}

In order to illustrate the equivalence of the continuum solution involving in the repulsive \Zstate\ of CH, we model a photo-absorption spectrum  \Xstate\ $\to$ \Cstate/\Zstate, where we follow the recipe from \citet{21PeYuTe} and \citet{23TePeZh}. For the \Xstate\ state, we use the same Morse function representation in Eq.~\eqref{e:Morse} with the parameters listed in Table~\ref{t:CH:params}.  For the transition electric dipole moments $\bar\mu_{X,C}$ = $\bra{\Xstate}\mu\ket{\Cstate}$ and $\bar\mu_{X,2}$ = $\bra{\Xstate}\mu\ket{\Zstate}$ of CH we adopt the \ai\ curves by \citet{87vanDishoeck} with an approximate model using the following function:
\begin{equation}
\label{e:bob}
\bar\mu(r)=  (c_0 + c_1 \xi_p) (1-\xi_p),
\end{equation}
where  $\xi_p$ is the \v{S}urkus \citep{84SuRaBo.method} variable given by:
\begin{equation}
\label{e:surkus}
\xi_p= \frac{r^{p}-r^{p}_{\rm ref}}{r^{p}+r^{p}_{\rm ref }}.
\end{equation}
The parameters defining the diabatic transition dipole moment (TDM) functions are listed in Table~\ref{t:CH:dipole}. The adiabatic TDM curves are obtained through the unitary transformation $U(r)$:  
\begin{equation}
\bm{\bar\mu}^{\rm a}(r) =\bm {\bar\mu}^{\rm d}(r){\bf U}^{\dagger}
= \left(\bar\mu^{\rm a}_1,\bar\mu^{\rm a}_2\right) 
=  \left(\bar\mu_1^{\rm d} \cos\beta - \bar\mu^{\rm d}_2 \sin\beta,   \bar\mu_1^{\rm d} \sin\beta + \bar\mu^{\rm d}_2 \cos\beta \right),
\label{eq:dia_V:2}
\end{equation}
where $\beta(r)$ is from Eq.~\eqref{eq:beta(r)} and $\bar\mu_1^{\rm d}$, $\bar\mu_2^{\rm d}$ are the diabatic TDM curves $\bra{\Xstate}\mu\ket{\Cstate}$ and $\bra{\Xstate}\mu\ket{\Zstate}$, respectively. 
The full photodissociation system, in both the adiabatic and diabatic representations, is illustrated in Fig.~\ref{f:CH:X-C-2}.

\begin{table}[htbp!]
\footnotesize
    \centering
    \caption{The molecular parameters defining the CH diabatic transition dipole moment functions}
    \label{t:CH:dipole}
    \begin{tabular}{lcccc}
        \hline
        Parameter &  $\bra{\Xstate}\mu\ket{\Cstate}$ &$\bra{\Xstate}\mu\ket{\Zstate}$ \\ 
        \hline\hline 
$r_{\rm ref}$, \AA    &  1.4   &  1.27    \\
$p$  & 4  & 5  \\ 
$c_0$, Debye  &  0.71 & 0.85  \\ 
$c_1$, Debye  &  0.09 & 0.17 \\ 
\hline
\end{tabular}
\end{table}

\begin{figure*}[ht!]
    \centering
    \includegraphics[width=.48\textwidth]{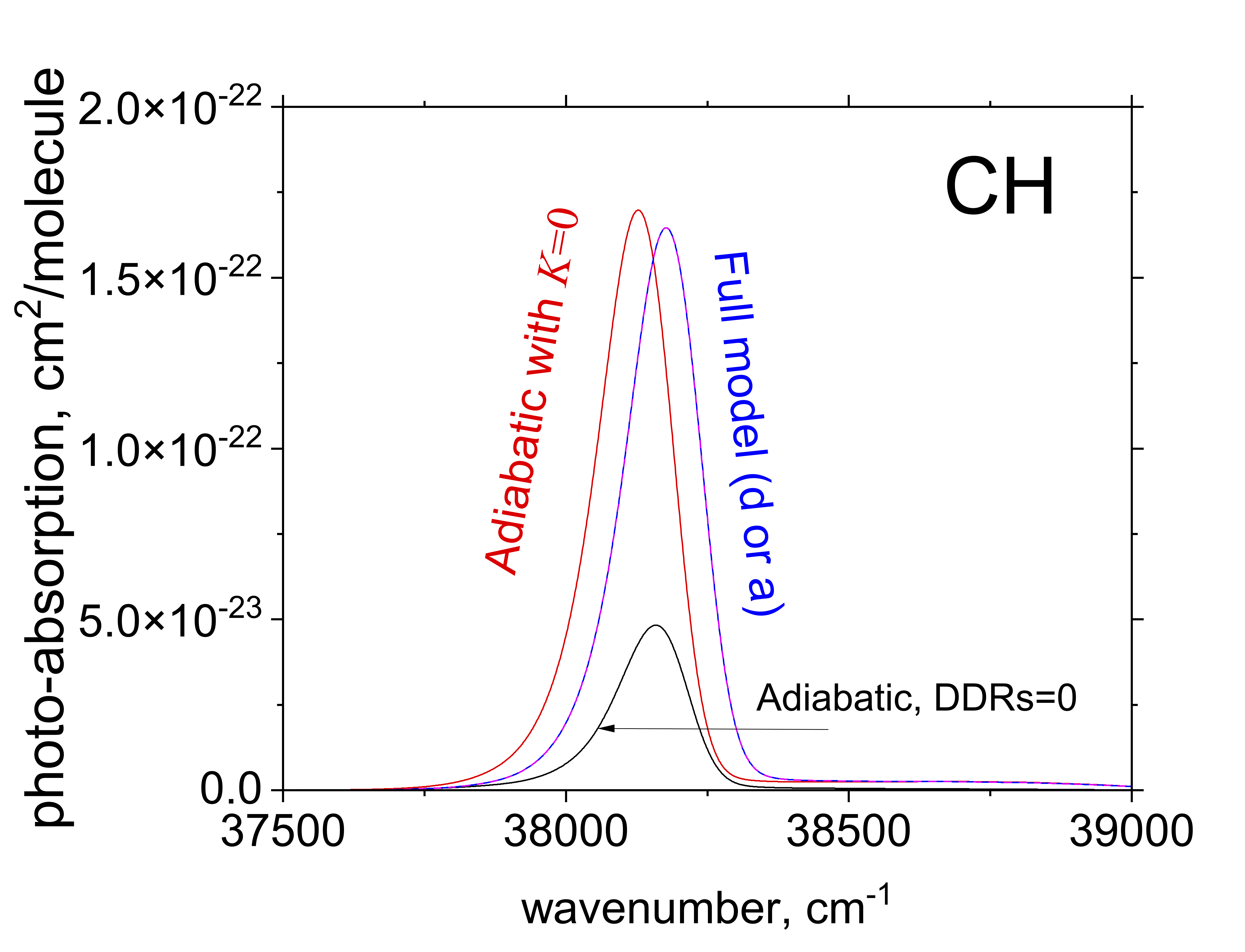}
    \caption{Photo-absorption spectra of CH at $T=300$~K. The no-approximation case is shown with the blue line; the NAC=0 case is shown with the red line and the black line shows spectrum with all DDRS set to zero.} 
    \label{f:ch:photo}
\end{figure*}

\begin{figure*}[ht!]
    \centering
    \includegraphics[width=.98\textwidth]{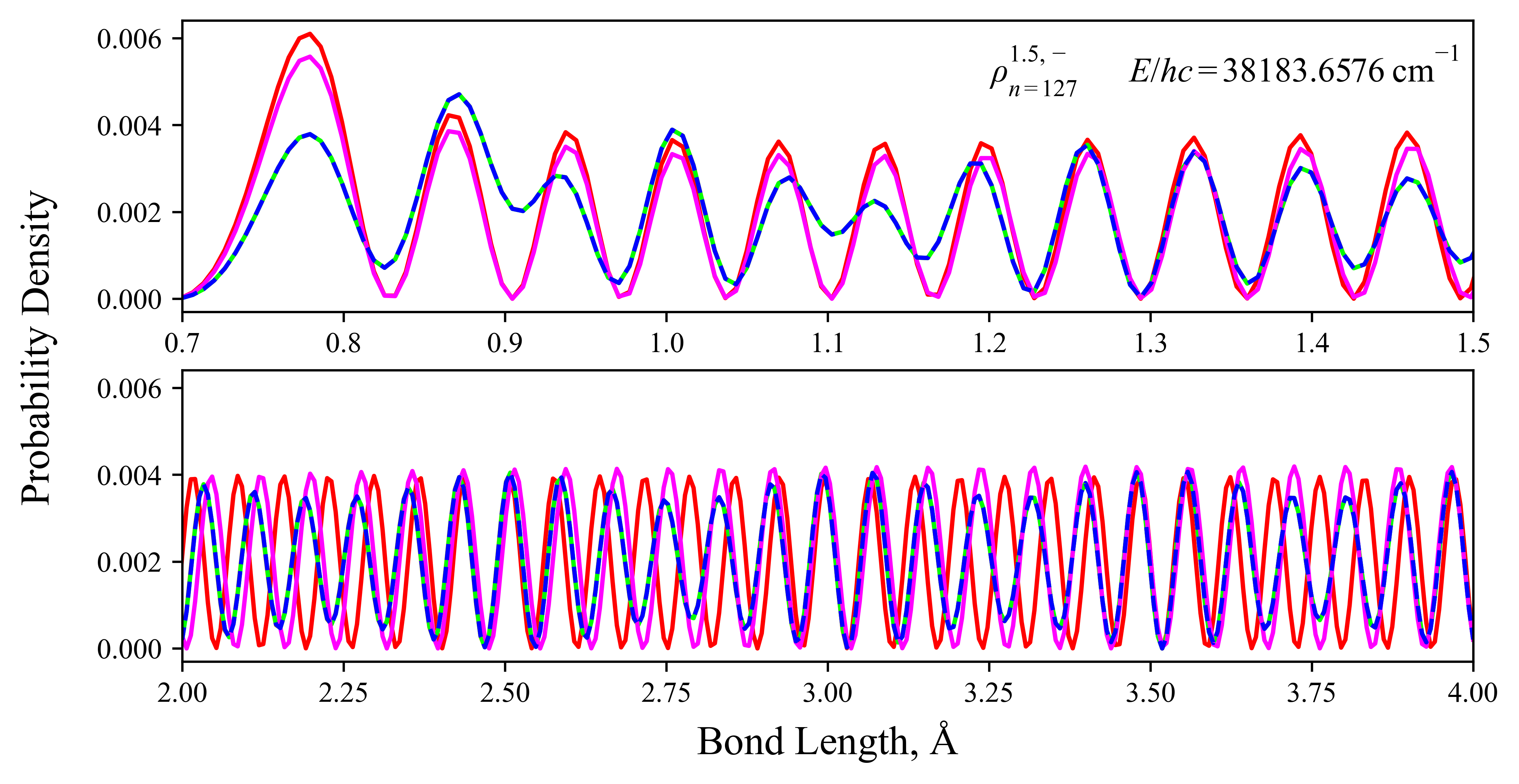}
\caption{Reduced density of the continuum state corresponding to an energy of $hc\cdot38183.6576$ \cm. Its transition with the \Xstate$(J=1.5,f,v=0)$ state is positioned at the peak in the spectra of Fig.(\ref{f:ch:photo}). The reduced density state is illustrated and computed using different levels of theory: diabatic representation with DC (blue dotted); diabatic model with the DC turned off (magenta, A3); adiabatic representation with both the NAC and $K$ correction included (lime green); adiabatic representation with NAC only (orange, A2); adiabatic representation with no correction (red, A1).} 
    \label{f:rho:CH_continuum}
\end{figure*}

Figure~\ref{f:ch:photo} shows a photo-absorption spectrum of CH at $T=300$~K computed with \Duo\ using the continuum solution of the coupled \Cstate/\Zstate\ system from the bound states of \Xstate, for the diabatic and adiabatic models. We used the box of 60~\AA\ and 1600 sinc-DVR points. For the cross sections, a Gaussian line profile of the half-width-at-half-maximum (HWHM) of 50~\cm\ was used to redistribute the absorption intensities between the `discrete' lines representing photo-absorption continuum. For details see \citet{21PeYuTe}. The diabatic and adiabatic continuum wavefunctions are obtained identical, so the photo-absorption spectra in this figure are indistinguishable.  Figure~\ref{f:ch:photo}  also illustrates  effects of the non-adiabatic  approximations on the photo-absorption spectra of CH. Removing the diagonal DDR ($K=0$) results in a shift of the band by about -50~\cm, while setting both DDRs to zero leads to a significant drop of the absorption by a factor of $\sim 4$. If we remove the DC term from the diabatic model, the bound absorption becomes dominating  in the Franck-Condon region (see Fig.~\ref{f:CH:X-C-2}) and the photo-absorption contribution drops by two order of magnitude and is therefore not visible on this scale. As a further illustration of the continuum system of CH, Fig.~\ref{f:rho:CH_continuum} gives an example of reduced densities of one of the continuum states used in the photo-absorption simulations.

\section{Convergence} \label{Convergence}

Since \Duo\ uses a solution of the $J=0$ uncoupled vibrational problem to form its vibrational basis set functions $\psi_v (r)$, and these model problems are hugely different depending on the representation, one can also expect that the convergence of the eigensolution to be impacted by the choice of the representation. 

\begin{figure}[ht!]
    \centering
    \includegraphics[width=.48\textwidth]{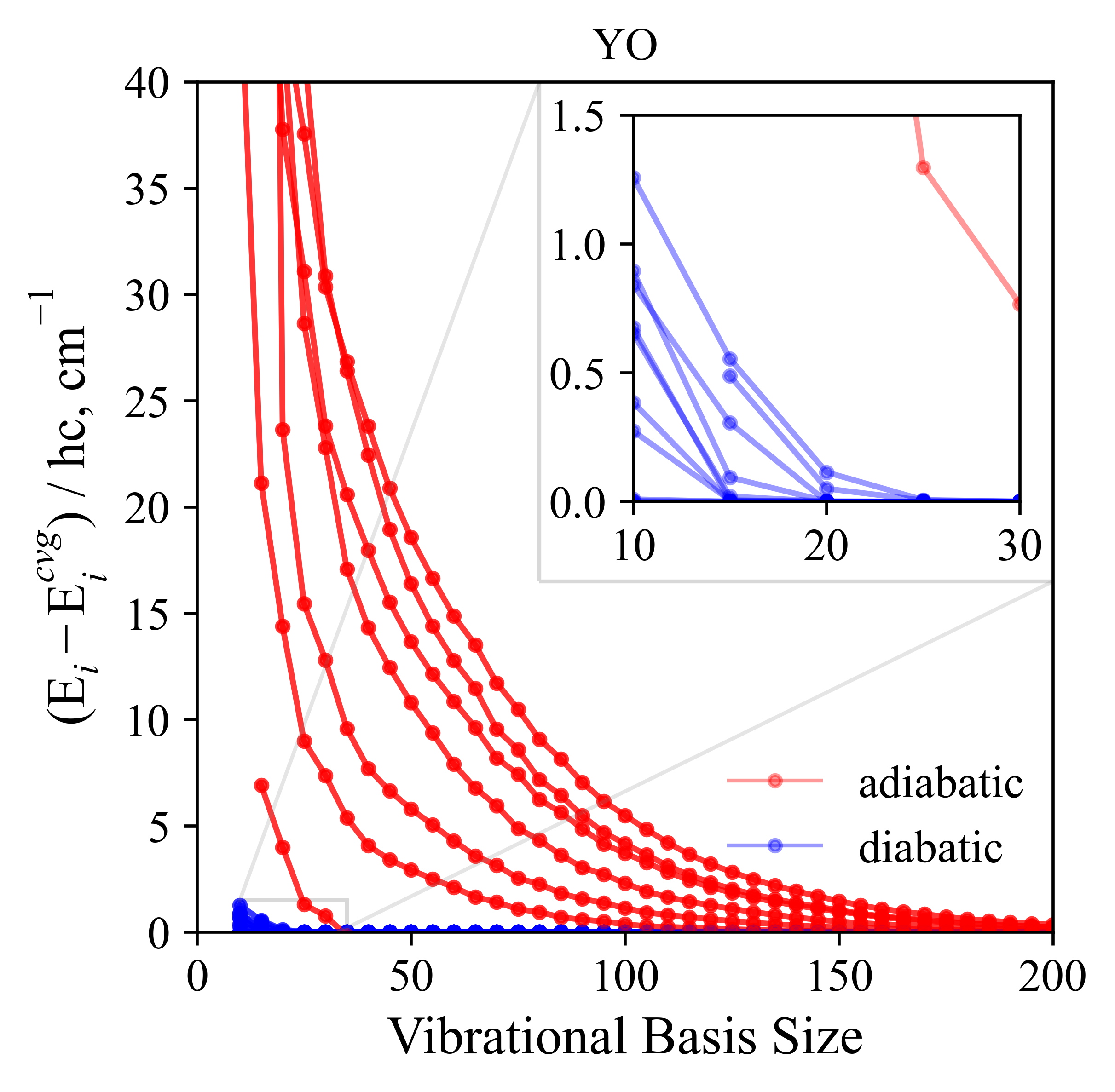}
    \includegraphics[width=.48\textwidth]{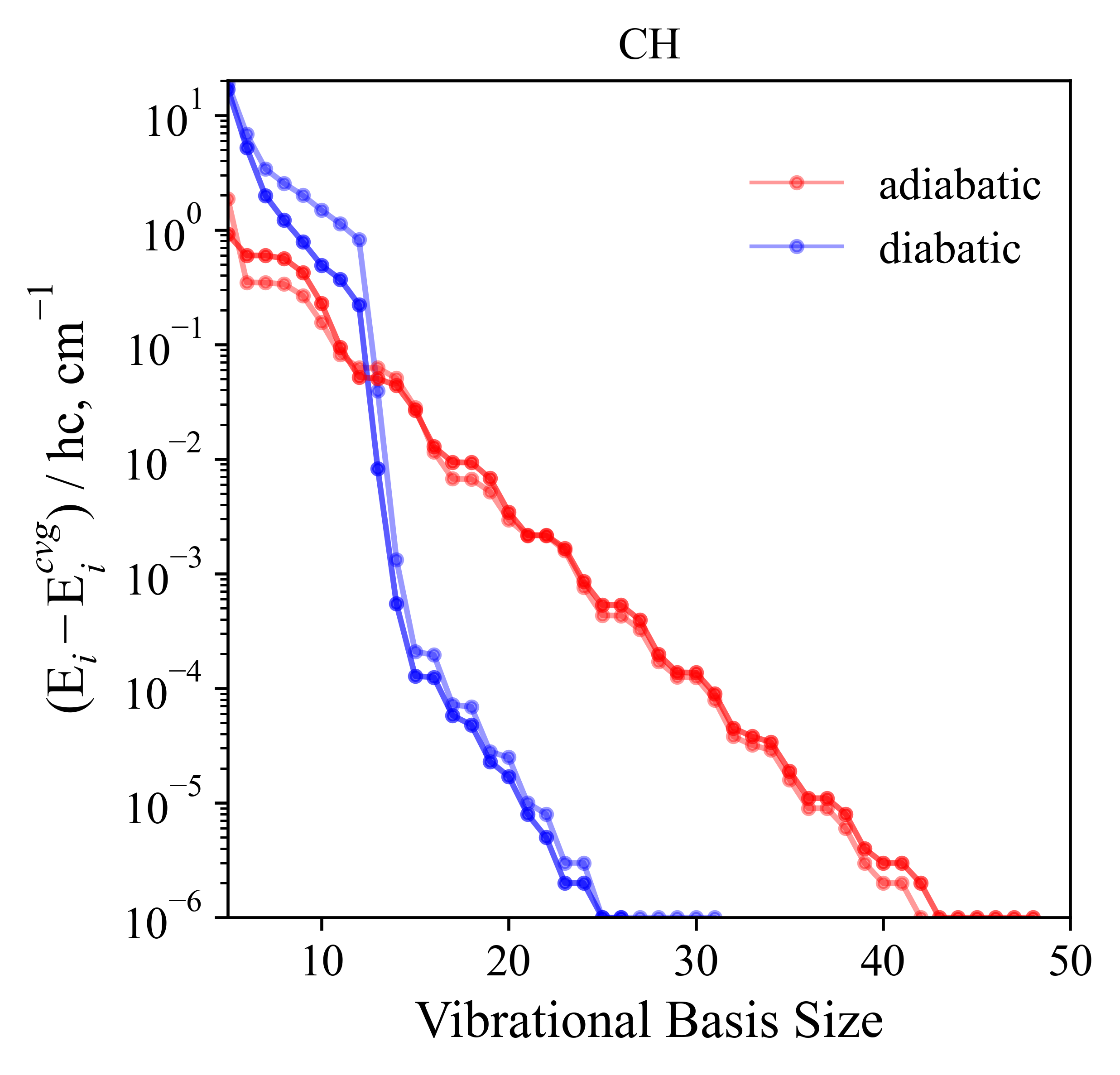}
    \caption{Convergence of the lowest 20 vibrational $J=0$ energies of the \DS\ state of YO (left) and the C~$^2\Sigma^+(v=0, e/f)$ state of CH (right) are plotted, where the difference of the $i$-th vibrational level $E_i$ to its converged value $E_i^{\rm cvg}$ is plotted as a function of vibrational basis size. A constant grid size of G$=3001,4001$ points for the sinc-DVR basis set was used for the YO and CH states, respectively. We see that the diabatically computed energies for YO converge much faster than the adiabatic ones, whereas for CH the opposite is true.} 
    \label{fig: convergence}
\end{figure}

Here we test the convergence of the $J=0.5$ energy levels of our simplified YO and CH models  in the diabatic and adiabatic representations  where all non-adiabatic effects  encountered. Figure  \ref{fig: convergence} illustrates the convergence of the lowest 20  $J=0.5$ energies of YO and the $n=5$ state of CH ($C^2\Sigma^+(J=0.5,\pm)$) where the difference of the $i$-th  level $\tilde{E}_i$ to its converged value $\tilde{E}_i^{\rm cvg}$ is plotted as a function of vibrational basis size. The two systems show contrasting results. The diabatically computed YO ($D^2\Sigma^+$) energies converge very quickly for basis sizes of $\sim 25$ whereas within the adiabatic representation a much larger basis set of $\sim 250$ was required to achieve convergence. For CH ($C^2\Sigma^+$) the adiabatic energies initially converge faster but the diabatic energies eventually converge to within $10^{-6}$ \cm\ for a basis size of $\sim25$ as opposed to $\sim42$ for the adiabatic energies.

Tests comparing the convergence rates for vibrational energies of higher $J$ resulted in the same conclusions as above for the $J=0.5$ case.

This shows that there is not one representation that rules over the other, it depends on the character of the avoided crossing, specifically in its position, the shape of the potentials approaching  the crossing, and the separation of the adiabatic PECs. It is therefore important to consider the system of study before choosing a representation, where all corrections must be included.

\section{Conclusion} \label{sec:conclusion}

A demonstration of the equivalency of the diabatic and adiabatic representations for two model diatomic systems, bound electronic \BS\ and \DS\  states of YO and abound/repulsive electronic system \Cstate\ and \Zstate\ of CH, is presented. Both representations should be equivalent by construction, but we explicitly show this within nuclear motion calculations through comparison of the rovibronic energies and wavefunctions. 
The importance of different non-adiabatic couplings in the molecular Hamiltonian are investigated, such as how the rovibronic energies and wavefunctions change when the NAC, the second derivative coupling or the diabatic coupling vanish.  

We present a transformation from the adiabatic to the strict diabatic basis for an isolated two-electronic state diatomic system. Each representation is defined by three functions, the adiabatic representation is given by two avoiding PECs and their corresponding NAC whereas the diabatic picture is analogously defined by two diabatic PECs and a DC, all of which are related to each other through the mixing angle. Because of this, any three of the aforementioned quantities can be used to fully reconstruct either the adiabatic or diabatic representation. We demonstrate that the choice of two diabatic PECs and a NAC provides an easily paramaterisable and powerful way to define the two-level problem. In the case of the diabatic PECs, they can be modelled easily by Morse oscillators, and the NAC is easily modelled using a Lorentzian.

We show that omission of any of the non-adiabatic terms lead to significant changes in the spectral properties of these systems, unsatisfactory especially for high-resolution applications. Even the diagonal derivative coupling, often omitted in practical applications, is shown to be of central importance when achieving equivalency. 

We also show that the choice of a preferable representation, diabatic or adiabatic, is not the same for all systems. For cases where the NAC is small (large DC) then the adiabatic representation shows fast convergence of rovibronic energy levels. However, for cases where the NAC is large (small DC) then the diabatic representation converges rovibronic energies with very small basis sets where large ones are required for the corresponding adiabatic representation.

We used simplified approximated functions to model different diabatic and adiabatic curves for the purpose of facilitating the comparison and demonstration of the equivalency, as well as to simplify the debugging process. In fact, our program \Duo\ uses numerically defined curves either provided as grids of $r$-dependent values or generated from analytic input functions as used here. For the convenience of the reader all curves from this work are provided both in the analytic and numerical representations as ASCII files, which are also  \Duo\ input files. As we  demonstrated,  the models provide exact equivalency of the diabatic and adiabatic salutations and therefore can be used as benchmark for similar programs. At the same time, \Duo\ provides an efficient platform to test  different aspects of diabatisations in diatomic calculations, including testing different approximations. \Duo\ is an open-access, with an extended online manual and many examples.

It would be interesting to develop and apply a similar methodology for polyatomic  molecules, where the derivative couplings cannot be fully transformed away. The exact equivalence of the two representations should be still possible to demonstrate numerically even for a quasi-diabatic transformation.  This work for triatomic molecules is currently underway.  In the present diatomic study we show that exclusion of the DDR couplings can lead to differences on the order of magnitude of 10s--100s of \cm\ in the energy wavenumbers, reinforcing the need for a careful error budget of all the approximations made when using them in high-resolution spectroscopic applications. 


\section{Associated Content}
All the DDR, potential energy, and DC curves are programmed in Duo analytically and are provided on a grid of 1000 equidistant bond lengths as part of the supplementary material. The spectroscopic models of CH and YO are also provided in the form of \Duo\ input files both in the diabatic and adiabatic representations as part of the supplementary material.

\section*{Author Information}

\subsection*{Corresponding Author}
\textbf{Sergei N.Yurchenko} - Department of Physics and Astronomy,
University College London, WC1E 6BT London, U.K.; 
https://orcid.org/0000-0001-9286-9501; Email: s.yurchenko@
ucl.ac.uk

\subsection*{Authors}

\textbf{Ryan P. Brady} - Department of Physics and Astronomy,
University College London, WC1E 6BT London, U.K.; 
https://orcid.org/
0000-0002-4161-5189

\noindent \textbf{Charlie Drury} - Department of Physics and Astronomy,
University College London, WC1E 6BT London, U.K.

\noindent \textbf{Jonathan Tennyson} - Department of Physics and Astronomy,
University College London, WC1E 6BT London, U.K.;
orcid.org/0000-0002-4994-5238

\subsection*{Notes}
There are no conflicts to declare.

\Duo\ is an open-source software, which is available at \href{https://github.com/Exomol/Duo}{https://github.com/Exomol/Duo}, where the DUO manual and many worked examples can be found.

\section*{Acknowledgements}

 This work was supported by UK STFC under grant ST/R000476/1. This work made  use of the STFC DiRAC HPC facility supported by BIS National E-infrastructure capital grant ST/J005673/1 and STFC grants ST/H008586/1 and ST/K00333X/1. We thank the European Research Council (ERC) under the European Union’s Horizon 2020  research and innovation programme through Advance Grant number  883830.

\subsection{Conditions for a Strictly Diabatic Basis: Nuclear Motion Part}
\label{subsec: Nuclear Motion Dia Transform}
Let us begin with the strict diabatic basis, a frame where both the off-diagonal DDR and DBOC couplings vanish simultaneously, which was shown to be possible for the diatom by \citet{82MeTrxx.ai}. Within this diabatic representation the Hamiltonian comprises of a purely diagonal kinetic energy operator and an electronic potential matrix with non-zero off-diagonal coupling elements analogous to the non-adiabatic couplings (DDRs) within the adiabatic representation, called diabatic couplings \citep{82MeTrxx.diabat,04JaKeMe.diabat} (DCs). All PECs are allowed to cross in this frame and property operator curves (e.g. SOCs) are smooth. We now wish to study the diabatic and adiabatic Hamiltonians in a similar vein to the work by \citet{84KoDoCe.diabat} who develop a Hamiltonian for the two-coupled electronic state problem. With this, the diabatic coupled-two-electronic state Born-Huang Hamiltonian takes the following matrix form
\begin{align*}
    \hat{H}^{(d)} = \hat{T}^{(d)} + \hat{V}^{(d)} \equiv -\frac{\hbar^2}{2\mu}\begin{pmatrix} \frac{d^2}{dr^2} & 0 \\ 0 & \frac{d^2}{dR^2}\end{pmatrix} + \begin{pmatrix} V^d_1 & V^d_{1,2} \\ V^d_{1,2} & V^d_2\end{pmatrix},
\end{align*}
where $V^d_{1,2}$ is the DC (potential) coupling the two electronic states in question, $\ket{\varphi^d_i}$ are the diabatic nuclear basis functions, $r$ is the diatom nuclear bond coordinate, and the superscripts denote the adiabatic 'a' or diabatic 'd' representation. To find the conditions required to achieve this strict diabatic basis, we transform back to the adiabatic representation via a inverse unitary transformation $\boldsymbol{U}^{\dagger}$ where $\boldsymbol{U}$ transforms the molecular system from the adiabatic to diabatic (AtDT) representation - the well known ATD transformation \citep{89Baer.diabat, 00Baer.diabat, 00BaAlxx.diabat,02BaerMichael.diabat, 06Baer, 82MeTrxx.diabat, 19YaXiZh,22ShVaZo.diabat} - via the following rotation
\begin{align*}
    U=\begin{pmatrix}
            \cos(\beta(r)) & -\sin(\beta(r)) \\ \sin(\beta(r)) & \cos(\beta(r))
          \end{pmatrix}.
\end{align*}
We define our transformation to the diabatic representation through the transform of the coupled PECs (see text) $\boldsymbol{V}^d = \boldsymbol{U}^{\dagger}\boldsymbol{V}^a\boldsymbol{U}$, or equivalently the change of basis via $\boldsymbol{U}=\sum_i\ket{\varphi^a_i}\bra{\varphi^d_i}$. Hence the diabatising transformation can be thought to either mix operator matrix elements or mix the adiabatic basis functions through a rotation of angle $\beta(r)$ via $\ket{\vec{\varphi}^{\;d}} = \boldsymbol{U^{\dagger}}\ket{\vec{\varphi}^{\;a}}$. This rotation is called the mixing angle \citep{99SiHaWe.diabat, 15AnBaxx.diabat, 17BaAnxx.diabat,18KaBeVa.diabat,22ShVaZo.diabat,19YaXiZh} and is a function of the diatom bond length. We can now transform the diabatic Hamiltonian to the adiabatic representation through action of $\boldsymbol{U}$ on the nuclear kinetic energy operator and then enclose with adiabatic nuclear basis to find the adiabatic matrix elements $T^{(a)}_{\alpha,\beta}$, we have in atomic units,
\begin{align*}
\bra{\varphi^a_\alpha}\Ut\hat{T}^{(d)}\Utdag\ket{\varphi^a_\beta} = -\braket{\varphi^a_\alpha |\Ut \frac{d^2}{dr^2}\Utdag|\varphi^a_\beta}.
\end{align*}
Before evaluating the above matrix elements, we introduce a minus sign and invert the direction of one of the first derivative components of the second derivative via a Hermitian conjugation which can be done under integration of nuclear coordinates. We show this explicitly via the following integration by parts in wavefunction notation,
\begin{align*}
  T^{(a)}_{\alpha,\beta} = -\int^{\infty}_{\-\infty}dr \; \varphi^{a*}_{\alpha}\Ut\frac{d^2}{dr^2}\Utdag\varphi^a_{\beta} = -\big[\varphi^{a*}_{\alpha}\Ut\frac{d}{dr}\Utdag\varphi^a_{\beta} \big]^{\infty}_{- \infty} +\int^{\infty}_{\-\infty}dr \; \frac{d(\varphi^{a*}_{\alpha}\Ut)}{dr}\frac{d(\Utdag\varphi^a_{\beta})}{dr}.
\end{align*}
Since the wavefunction vanishes for $|r|\rightarrow\infty$  then we have $\big[\varphi^{a*}_{\alpha}\Ut\frac{d}{dr}\Utdag\varphi^a_{\beta} \big]^{\infty}_{- \infty}= 0$. Thus, we are left with
\begin{align*}
 T^{(a)}_{\alpha,\beta} = \int^{\infty}_{\-\infty}dr \; \frac{d(\varphi^{a*}_{\alpha}\Ut)}{dr}\frac{d(\Utdag\varphi^a_{\beta})}{dr} \equiv \derivBraket{\varphi^a_{\alpha}\Ut}{\Utdag\varphi^a_\beta}.
\end{align*}
We will see that this inversion of the direction of derivatives will ultimately lead to the adiabatic Hamiltonian being in its Hermitian form and will be convenient for us to find the conditions of a strictly diabatic basis. Evaluating the derivatives using the product rule and expanding the equation we find the following matrix elements,
\begin{align*}
 T^{(a)}_{\alpha,\beta} = -\braket{\varphi^a_\alpha |\Ut \frac{d^2}{dr^2}\Utdag|\varphi^a_\beta} = \derivBraket{\varphi^a_{\alpha}\Ut}{\Utdag\varphi^a_\beta} = \left[ \derivBra{\varphi^a_\alpha}\Ut + \bra{\varphi^a_\alpha}\frac{d\Ut}{dr}\right] \left[ \frac{d \Utdag}{dr}\ket{\varphi^a_\beta} + \Utdag\derivKet{\varphi^a_\beta}\right] \\
= \derivBra{\varphi^a_\alpha}\Ut\frac{d \Utdag}{dr}\ket{\varphi^a_\beta} + \derivBra{\varphi^a_\alpha}\Ut\Utdag\derivKet{\varphi^a_\beta} +  \bra{\varphi^a_\alpha}\frac{d\Ut}{dr} \frac{d \Utdag}{dr}\ket{\varphi^a_\beta} + \bra{\varphi^a_\alpha}\frac{d\Ut}{dr}\Utdag\derivKet{\varphi^a_\beta}
\end{align*}
We now want to write this in a convenient form whereby the diagonal and off-diagonal terms are recognised. To this end, consider the derivative matrix product $\frac{d\Ut\Utdag}{dr}$
\begin{align*}
\frac{d\Ut\Utdag}{dr} = \frac{d\Ut}{dr}\Utdag + \Ut\frac{d\Utdag}{dr} = 0,
\end{align*}
since $\Ut\Utdag$ is the identity. This infers that the products $\frac{d\Ut}{dr}\Utdag$ and $\Ut\frac{d\Utdag}{dr}$ are equal to a skew-symmetric matrix since $S+S^{\dagger}=0$. Hence,
\begin{align*}
\frac{d\Ut}{dr}\Utdag &= S \rightarrow \frac{d\Ut}{dr} = SU \\
\Ut\frac{d\Utdag}{dr} &= S^{\dagger} \rightarrow \frac{d\Utdag}{dr} = \Utdag S^{\dagger}.
\end{align*}
We can now see that the derivative matrix product $\frac{d\Ut}{dr}\frac{d\Utdag}{dr}$ is given in terms of the skew-symmetric matrix via
\begin{align*}
\frac{d\Ut}{dr}\frac{d\Utdag}{dr} = S\Ut\Utdag S^{\dagger} = -S^2.
\end{align*}
For $2\times2$ unitary matrices, the corresponding skew symmetric matrix $S$ only has one unique element, $\zeta$, and the squared matrix $S^2$ has every diagonal element equal to  $-\zeta^2$. Thus, for a 2-electronic state problem, the unitary matrix product $\frac{d\Ut}{dr}\frac{d\Utdag}{dr}$ is diagonal. Through a similar argument, the matrix product $\Ut\frac{d\Utdag}{dr}$ is a skew-symmetric matrix, with zero in the diagonal. We can thence write the kinetic energy matrix element $T^{(a)}_{\alpha,\beta}$ as
\begin{align*}
 T^{(a)}_{\alpha,\beta}
= \bra{\varphi^a_\alpha}\left(\overleftarrow{\frac{d}{dr}}\textbf{I}\overrightarrow{\frac{d}{dr}}-\textbf{S}^2\right)\ket{\varphi^a_\beta} 
+ \bra{\varphi^a_\alpha}\left[ \overleftarrow{\frac{d}{dr}}\Ut\frac{d\Utdag}{dr} + \frac{d\Ut}{dr}\Utdag\overrightarrow{\frac{d}{dr}} \right]\ket{\varphi^a_\beta}.
\end{align*}
The first term contains the diagonal matrix elements and the second term are the off-diagonal terms. As we will see in the next section, $\Ut\frac{d\Utdag}{dr}$ is equivalent to the first-order NAC matrix, $W^{(1)}$, and defines the condition to make this NAC vanish upon action of the AtDT. The first-order matrix differential equation $\Ut\frac{d\Utdag}{dr}=W^{(1)}$ has been investigated thoroughly by Baer and co-authors since the late 1980s \citep{89Baer.diabat, 00Baer.diabat, 00BaAlxx.diabat,02BaerMichael.diabat, 06Baer}. These works develop the so called line-integral approach in solution to the above matrix differential equation, which for a two-level system, is easily achieved by noticing the common solution to the first order ODE $U=e^{-\int{W^{(1)}dr}}$. It turns out this solution is exact for the diatomic two-coupled electronic state problem and arises because $W^{(1)}$ and its integral commute. One can then manipulate the taylor series expansion for $U=e^{-\int{W^{(1)}dr}}$ to show that it is indeed equal to a rotation matrix of the mixing angle $\beta$. To this end, let us now insert the rotation matrix form for $\Ut$ into the above expression for the kinetic energy matrix elements, noting the following matrix multiplications
\begin{align*}
    \Ut\Utdag &=     \begin{pmatrix}
        1 & 0 \\ 0 & 1
    \end{pmatrix}=\textbf{I},\\
    \Ut\frac{d\Utdag}{dr} &= 
    \begin{pmatrix}
        0 & 1 \\ -1 & 0 
    \end{pmatrix}
    \frac{d \beta(r)}{dr} = \tilde{\boldsymbol{\sigma}} \frac{d \beta(r)}{dr}, \\
    \frac{d\Ut}{dr} \Utdag& = 
    \begin{pmatrix}
        0 & -1 \\ 1 & 0 
    \end{pmatrix}
    \frac{d \beta(r)}{dr} = -\tilde{\boldsymbol{\sigma}} \frac{d \beta(r)}{dr}, \\
    \frac{d\Ut}{dr}\frac{d\Utdag}{dr} &= \textbf{I}
    \left(\frac{d \beta(r)}{dr}\right)^2, \\
\end{align*}
where $\tilde{\boldsymbol{\sigma}} = i \boldsymbol{\sigma}$ and $\boldsymbol{\sigma}$ is the second Pauli matrix. We can now simplify our expression for the adiabatic kinetic energy matrix elements by use of the above relations, giving us
\begin{align*}
-\braket{\varphi^a_\alpha |\Ut \frac{d^2}{dr^2}\Utdag|\varphi^a_\beta}
= \derivBra{\varphi^a_\alpha}\tilde{\boldsymbol{\sigma}}\frac{d \beta}{dr}\ket{\varphi^a_\beta} 
+ \derivBra{\varphi^a_\alpha}\textbf{I}\derivKet{\varphi^a_\beta} 
+ \bra{\varphi^a_\alpha}\textbf{I}(\frac{d \beta}{dr})^2\ket{\varphi^a_\beta}
-\bra{\varphi^a_\alpha}\tilde{\boldsymbol{\sigma}}\frac{d\beta}{dr}\derivKet{\varphi^a_\beta}\\
= \derivBra{\varphi^a_\alpha}\textbf{I}\derivKet{\varphi^a_\beta} 
+ \bra{\varphi^a_\alpha}\textbf{I}\left(\frac{d \beta}{dr}\right)^2\ket{\varphi^a_\beta}
+ \bra{\varphi^a_\alpha}\left[ \overleftarrow{\frac{d}{dr}}\tilde{\boldsymbol{\sigma}}\frac{d \beta}{dr} - \tilde{\boldsymbol{\sigma}}\frac{d\beta}{dr}\overrightarrow{\frac{d}{dr}} \right]\ket{\varphi^a_\beta}
\end{align*}
Noting that the adiabatising transformation diagonalises the electronic Hamiltonian by construction, performing a Hermitian conjugation of one of the first derivative components in the second term above under integration of the nuclear coordinated to write $\derivBra{\varphi^a_\alpha}\textbf{I}\derivKet{\varphi^a_\beta} = -\braket{\varphi^a_\alpha | \textbf{I} \frac{d^2}{dr^2} | \varphi^a_\beta}$, we can write the adiabatic 2$\times$2 Born-Huang Hamiltonian as, going back to SI units by introduction of the $\hbar^2/2\mu$ kinetic energy factor,
\begin{align}
\label{eq:adi_H_matel}
    \hat{H}^{(a)} =  -\frac{\hbar^2}{2\mu}\begin{pmatrix}\frac{d^2}{dr^2} -\left( \frac{d\beta}{dr}\right)^2 &-\left[ \overleftarrow{\frac{d}{dr}}\frac{d \beta}{dr} -\frac{d\beta}{dr}\overrightarrow{\frac{d}{dr}} \right] \\\left[ \overleftarrow{\frac{d}{dr}}\frac{d \beta}{dr} -\frac{d\beta}{dr}\overrightarrow{\frac{d}{dr}} \right] & \frac{d^2}{dr^2} - \left( \frac{d\beta}{dr}\right)^2\end{pmatrix} + \begin{pmatrix} V^a_1 & 0 \\ 0 & V^a_2\end{pmatrix}.
\end{align}
At the moment is not clear if it is Hermitian (see next section) and off-diagonal terms to the nuclear kinetic energy operator are introduced, which will correspond to the non-adiabatic derivative couplings, or DDRs, in this representation. Here the direction of the derivatives are still shown, and is how we implement the matrix elements within our nuclear motion code \duo. We can again introduce a minus sign and redirect the derivative through Hermitian conjugate to arrive at the following
\begin{align}
    \hat{H}^{(a)} =  -\frac{\hbar^2}{2\mu}\begin{pmatrix} \frac{d^2}{dr^2} - \left( \frac{d\beta}{dr}\right)^2 & -\left[ -\frac{d}{dr}\frac{d \beta}{dr} -\frac{d\beta}{dr}\frac{d}{dr} \right] \\ \left[ -\frac{d}{dr}\frac{d \beta}{dr} -\frac{d\beta}{dr}\frac{d}{dr} \right] &\frac{d^2}{dr^2} - \left( \frac{d\beta}{dr}\right)^2\end{pmatrix} + \begin{pmatrix} V^a_1 & 0 \\ 0 & V^a_2\end{pmatrix},
\label{eq:adi_H}
\end{align}
Since at this point the electronic DoF have been already integrated over, we can identify the scalar terms in the above Hamiltonian with the DDRs one would get from \ai\ calculations. The only scalar term that arises is the derivative of the mixing angle, and therefore the condition to transform to the strict diabatic basis depends solely on the transformation through  $\frac{d \beta}{dr}$, the so called Abelian case as discussed by \citet{06Baer}. We know that the DDRs will consist of a first DDR, a.k.a. the NAC, which we call generally $W^{(1)}$ here which is antihermitian, and a second DDR diagonal coupling element which we call $K$. Thus, the condidtions for a strictly diabatic basis can be summarised by
\begin{align*}
    \hat{W}^{(1)} &= \begin{pmatrix}
        0 & \frac{d\beta}{dr} \\
        -\frac{d\beta}{dr} & 0
    \end{pmatrix} ,\\
    \hat{K} &= \begin{pmatrix}
        \left(\frac{d\beta}{dr}\right)^2 & 0\\
         0 & \left(\frac{d\beta}{dr}\right)^2
    \end{pmatrix} ,\\
\end{align*}
which we will show in the next section to be a sensible set of condiditions, thus providing a way to form the strict diabatic basis for diatomic molecules in the isolated two-electronic state problem.

\subsection{The Non-Adiabatic Derivative Couplings: Electronic Structure Part}
We now turn to the electronic structure part of our discussion on adiabatic and diabatic representations for the diatomic molecular Hamiltonian. Let us begin with the electronic matrix elements of the nuclear kinetic energy operator, or second derivative with respect to the nuclear bond length
\begin{align*}
    W^{(2)}_{\alpha,\beta} = \bra{\psi^a_\alpha}\frac{d^2}{dr^2}\ket{\psi^a_\beta},
\end{align*}
where $\ket{\psi^a}$ are the adiabatic electronic basis wavefunctions and $W^{(2)}_{\alpha,\beta}$ are the second DDR derivative coupling elements. By evaluating these matrix elements we are relaxing the Born-Oppenheimer approximation which would assume these elements to be zero. The NAC, or 1\ts{st} DDR, is well known and defined as the matrix elements of the first derivative
\begin{align*}
    W^{(1)}_{\alpha,\beta} = \begin{cases}
   \bra{\psi^a_\alpha}\frac{d}{dr}\ket{\psi^a_\beta}=  -\bra{\psi^a_\beta}\frac{d}{dr}\ket{\psi^a_\alpha}  & \text{if } \alpha\neq\beta\ \\
   0     & \text{if } \alpha = \beta \\
\end{cases},
\end{align*}
where the NAC is anti-Hermitian and can be investigated using the Hellman-Fynman theorum. To this end, consider the following eigenvalue equation
\begin{align*}
\hat{H}^{(a)}\ket{\psi^a_{\beta}} = E_\beta\ket{\psi^a_\beta},
\end{align*}
where $E_\beta$ are the electronic potential energies. Let us enclose by a bra state $\bra{\psi^a_\alpha}$
\begin{align*}
\bra{\psi^a_\alpha}\hat{H}^{(a)}\ket{\psi^a_\beta} = E_\beta \langle\psi^a_\alpha|\psi^a_\beta\rangle  = E_\beta\delta_{\alpha,\beta},
\end{align*}
where $\delta_{\alpha,\beta}$ is the Kroncker delta since $\ket{\psi^a}$ form an orthonormal basis. Considering the off-diagonal terms, or Born-Oppenheimer corrections, and performing the first derivative in bond length
\begin{align*}
\frac{d}{dr}\bra{\psi^a_\alpha}\hat{H}^{(a)}\ket{\psi^a_{\beta}} = 0 \\
\rightarrow  \bigBrktOp{\frac{d\psi^a_\alpha}{dr}}{\hat{H}^{(a)}}{\psi^a_{\beta}} + \bigBrktOp{\psi^a_\alpha}{\frac{d\hat{H}^{(a)}}{dr}}{\psi^a_{\beta}} + \bigBrktOp{\psi^a_\alpha}{\hat{H}^{(a)}}{\frac{d\psi^a_{\beta}}{dr}}  \\ = E_\beta\bigBrkt{\frac{d\psi^a_\alpha}{dr}}{\psi^a_{\beta}} + \bigBrktOp{\psi^a_\alpha}{\frac{d\hat{H}^{(a)}}{dr}}{\psi^a_{\beta}} + E_\alpha\bigBrkt{\psi^a_\alpha}{\frac{d\psi^a_{\beta}}{dr}} = 0.
\end{align*}
Where in the last part we used the Hermiticity of the Hamiltonian to act $\hat{H}^{(a)}$ on the bra yielding us $E_\alpha$. We now have,
\begin{align}
\label{eq:Hellmann-Feynmann NAC}
(E_\alpha-E_\beta)W^{(1)}_{\alpha,\beta}+ \bigBrktOp{\psi^a_\alpha}{\frac{d\hat{H}^{(a)}}{dr}}{\psi^a_{\beta}} = 0 \notag\\
\rightarrow W^{(1)}_{\alpha,\beta} = \frac{1}{E_\beta-E_\alpha}\bigBrktOp{\psi^a_\alpha}{\frac{d\hat{H}^{(a)}}{dr}}{\psi^a_{\beta}}.
\end{align}
Hence, if states $\alpha$ and $\beta$ are sufficiently well seperated, i.e. $|E_\beta-E_\alpha| \gg 1$, then the DDR matrix elements are small $W^{(1)}_{\alpha,\beta}\ll1$.
Multiple studies \citep{86LeYaxx.diabat,87SaYaxx.diabat, 06Baer} have used the following identity to relate $W^{(2)}$ and $W^{(1)}$ via
\begin{align}
\label{eq:W2=dW1-K}
   \frac{d}{dr}\braket{\psi^a_\alpha | \frac{d}{dr} | \psi^a_\beta} = \derivBraket{\psi^a_\alpha}{\psi^a_\beta} + \bra{\psi^a_\alpha}\frac{d^2}{dr^2}\ket{\psi^a_\beta} \rightarrow W^{(2)}_{\alpha,\beta} = \frac{d W^{(1)}_{\alpha,\beta}}{dr}-K_{\alpha,\beta},
\end{align}
where the DBOC $K_{\alpha,\beta}=\derivBraket{\psi^a_{\alpha}}{\psi^a_\beta}\neq 0$ for $\alpha=\beta$. Historically, the use of the above identity was to avoid the cumbersome computation of $W^{(2)}$ by only having to evaluate $W^{(1)}$ and its derivative, and is identical to the g-, h-, and k- notation by \citet{86LeYaxx.diabat,87SaYaxx.diabat} with $h = \frac{dg}{dr}-k$. The above relation shows the second DDR to have both diagonal and off-diagonal components and can be compared to the scalar terms in Eq.(\ref{eq:adi_H}). We can simplify further our expressions for the DDRs by introducing a resolution of the identity between the bra and ket of the DBOC $K_{\rho,\rho}$ within the adiabatic basis, yielding
\begin{align}
\label{eq:NAC_square_hellman-fynman}
   K_{\rho,\rho} &= \derivBraket{\psi^a_\rho}{\psi^a_\rho}
   = \sum^N_\kappa \bigBrkt{\frac{d \psi^a_\rho}{dr}}{\psi^a_\kappa}\bigBrkt{\psi^a_\kappa}{\frac{d \psi^a_\rho}{dr}}
   = \sum^N_\kappa W^{(1)}_{\rho,\kappa}W^{(1)}_{\kappa,\rho} \notag\\&= \sum^N_\kappa \frac{1}{(E_\rho-E_\kappa)(E_\kappa-E_\rho)}\bigBrktOp{\psi^a_\rho}{\frac{d\hat{H}^{(a)}}{dr}}{\psi^a_{\kappa}}\bigBrktOp{\psi^a_\kappa}{\frac{d\hat{H}^{(a)}}{dr}}{\psi^a_{\rho}}
\end{align}
where the summation is over all adiabatic states and in the last line we inserted the Hellman-Feynman relation in Eq.(\ref{eq:Hellmann-Feynmann NAC}). We see that for the coupled two-electronic state system, states $\ket{\psi^a_1}$ and $\ket{\psi^a_2}$, the NAC elements coupling  other states will be reduced by a factor of $\frac{1}{(E_\rho-E_\kappa)(E_\kappa-E_\rho)}$, for sufficiently well separated states from the coupled system $\ket{\psi^a_1}$ and $\ket{\psi^a_2}$ we can truncate the summation to over these two states only. We now have
\begin{align}
\label{eq:NAC_square}
   K_{\rho,\rho} \approx -\left(W^{(1)}_{1,2}\right)^2 
\end{align}

which is diagonal and equal to minus the NAC squared. We now see that all DDRs are completely defined in terms of the 1\ts{st} DDR, or NAC, only. Knowing we will want to find matrix elements within the nuclear basis later, we rewrite Eq.(\ref{eq:W2=dW1-K}) in matrix form 
\begin{align}
\label{eq:W2=dW1-W1^2}
  W^{(2)}_{\alpha,\beta} \approx \frac{d W^{(1)}_{\alpha,\beta}}{dr}+\left( W^{(1)}_{1,2}\right)^2\delta_{\alpha,\beta}.
\end{align}
In preperation to finding the matrix elements of the kinetic energy operator in nuclear basis, one would left and right multiply $\hat{T}$ by the Born-Huang wavefunction $\ket{\Psi_{tot}^a}=\ket{\psi^a}\ket{\varphi^a}$. However, let us consider only the electronic part and use the relations we have derived in this section, we have
\begin{align*}
     \braket{\psi^a_\alpha | \hat{T}^{(a)} | \psi^a_\beta} = -\frac{\hbar^2}{2\mu}\bra{\psi^a_\alpha}\overrightarrow{\frac{d^2}{dr^2}}\ket{\psi^a_\beta} = -\frac{\hbar^2}{2\mu}W^{(2)}_{\alpha,\beta} = -\frac{\hbar^2}{2\mu}\left(\overrightarrow{\frac{d}{dr}} W^{(1)}_{\alpha,\beta}+\left( W^{(1)}_{1,2}\right)^2\delta_{\alpha,\beta}\right).
\end{align*}
Where we have explicitly shown the direction of the derivative. Knowing we will take matrix elements within the nuclear basis we now push through the first derivative on the first-order NAC to write $ \overrightarrow{\frac{d}{dr}} W^{(1)}_{\alpha,\beta} =\frac{d W^{(1)}_{\alpha,\beta}}{dr} + W^{(1)}_{\alpha,\beta} \overrightarrow{\frac{d}{dr}}$. Under the integration of nuclear DoF we can also perform a Hermitian conjugation on the $\overrightarrow{\frac{d}{dr}} W^{(1)}_{\alpha,\beta}$ term to write $\overrightarrow{\frac{d}{dr}} W^{(1)}_{\alpha,\beta}=-\overleftarrow{\frac{d}{dr}} W^{(1)}_{\alpha,\beta}$. With this, we now have
\begin{align*}
     \braket{\psi^a_\alpha | \hat{T}^{(a)} | \psi^a_\beta} = -\frac{\hbar^2}{2\mu}\left(-\left[ \overleftarrow{\frac{d}{dr}}W^{(1)}_{\alpha,\beta}- W^{(1)}_{\alpha,\beta}\overrightarrow{\frac{d}{dr}}\right]+\left( W^{(1)}_{1,2}\right)^2\textbf{I}\right).
\end{align*}
The associated coupled-two-electronic state Born-Huang adiabatic correction Hamiltonian reads as
\begin{align*}
-\frac{\hbar^2}{2\mu}\begin{pmatrix} \left(W^{(1)}_{1,2}\right)^2 & -\left[ \overleftarrow{\frac{d}{dr}}W^{(1)}_{1,2} -W^{(1)}_{1,2}\overrightarrow{\frac{d}{dr} }\right] \\ \left[ \overleftarrow{\frac{d}{dr}}W^{(1)}_{1,2} -W^{(1)}_{1,2}\overrightarrow{\frac{d}{dr} } \right] & \left( W^{(1)}_{1,2}\right)^2\end{pmatrix} ,
\end{align*}
which is Hermitian and we used the antihermiticity of $W^{(1)}_{\alpha,\beta}$ to introduce another minus sign on the element $\braket{\psi^a_2 | \hat{T}^{(a)} | \psi^a_1}$. Comparison with the adiabatic molecular Hamiltonian in Eq.(\ref{eq:adi_H_matel}) yields the final conditions for transforming to a strictly diabatic basis to be
\begin{align}
\label{eq:dBeta/dr}
    W^{(1)}_{1,2} &= \brkteq{\psi^a_1}{\frac{d}{dr}}{\psi^a_2} = \frac{d\beta(r)}{dr},\\
    \derivBraket{\psi^a_\kappa}{\psi^a_\kappa} &= \left(W^{(1)}_{1,2}\right)^2 = \left( \frac{d\beta}{dr}\right)^2.
\end{align}
The condition that the 2\ts{nd} DDR must equal the square of the NAC in order to completely remove the DDR couplings is then sensible for a coupled two-electronic state system which is energetically well-separated from other adiabatic states, as shown in Eq.(\ref{eq:NAC_square_hellman-fynman}) under Hellman-Fynman formalism. This of course only works when the two-level system is effectively isolated, and breaks down upon approach of closely lying adiabatic states with common symmetries.

\bibliography{./bib/journals_phys,./bib/diabatisation,./bib/programs,./bib/sy,./bib/YO,./bib/methods,./bib/diatomic,./bib/abinitio,./bib/Books,./bib/CH,./bib/NaCl,./bib/BF++,./bib/stars,./bib/C2,./bib/HCl,./bib/jtj,./bib/H3+,./bib/H2} 

%

\end{document}